\documentclass[twocolumn]{aastex631}

\usepackage{amsmath}
\usepackage{multirow}
\usepackage{xcolor}
\usepackage{hyperref}

\newcommand{\usedinput}[1]{{\bfseries\boldmath #1}}

\begin{document}

\title{\Large Gaia neutron stars: demographics, birth rates, and connections to other neutron star populations}

\author[0000-0003-1247-9349]{Cheyanne Shariat}
\affiliation{Department of Astronomy, California Institute of Technology, 1200 E. California Blvd, Pasadena, CA, 91125, USA}

\author[0000-0001-6970-1014]{Natsuko Yamaguchi}
\affiliation{Department of Astronomy, California Institute of Technology, 1200 E. California Blvd, Pasadena, CA, 91125, USA}

\author[0000-0002-6871-1752]{Kareem El-Badry}
\affiliation{Department of Astronomy, California Institute of Technology, 1200 E. California Blvd, Pasadena, CA, 91125, USA}

\author[0000-0003-0976-4755]{Thomas J. Maccarone}
\affiliation{Department of Physics \& Astronomy, Box 41051, Texas Tech University, Lubbock TX 79409-1051, USA}

\begin{abstract}
Gaia DR3 revealed a new class of neutron stars (NSs) in au-scale orbits with solar-type companions. How these `Gaia NSs' form remains unclear, and constraining their intrinsic population properties is a necessary step toward understanding their origin. However, the current sample of 27 Gaia NS candidates is strongly biased by Gaia's astrometric sensitivity. In this work, we carefully model the Gaia selection function and incorporate it into a hierarchical Bayesian framework to infer the \textit{intrinsic} demographics of the Gaia NS population. We find that the population favors a decreasing period distribution with power-law slope $\alpha_P=-0.75 \pm 0.35$, a visible-star mass distribution peaked near $M_1\approx 0.85~{\rm M_\odot}$ ($\sigma_{M_1} \approx 0.2~{\rm M_\odot}$), an NS mass distribution peaked near $M_{\rm NS}\approx 1.3~{\rm M_\odot}$ with a tail towards higher masses, and an eccentricity distribution weighted toward moderate-to-large eccentricities. The intrinsic space density is $n\approx120~{\rm kpc^{-3}}$ within 2 kpc, implying that only $\sim3\%$ of the local Gaia NS population has been discovered so far.  Our forecasts predict that Gaia DR4 will increase the sample to $\sim300$ -- 650 NS candidates, including access to longer orbital periods that will provide direct tests of formation models. Finally, we place Gaia NSs in the context of the broader Galactic NS population by comparing their intrinsic birth rate, lifetime, and size to other classes of isolated and binary NSs. Only 0.01\% -- 1\% of all NSs are born into au-scale orbits with solar-type companions, making the birth rate of Gaia NSs 10 -- 100 times smaller than that of high-mass X-ray binaries and broadly comparable to symbiotic X-ray binaries. 
\end{abstract}


\section{Introduction}\label{sec:introduction}

Neutron stars (NSs) are the remnants of most massive stars ($\gtrsim8~{\rm M_\odot}$) after they undergo supernovae. $80$ -- $100\%$ of NS progenitors, massive OB-type stars, are born in binary or higher-order multiples \citep[][]{Sana2012,Moe2017}, yet most ($\gtrsim99\%$) young Galactic radio pulsars are isolated \citep{Lorimer2008}. This contrast implies that over the course of binary evolution and NS formation, most massive stars merge or become unbound from their companions.

Indeed, the space velocities of isolated pulsars support the notion that most NSs are born with large recoil kicks, often hundreds of ${\rm km~s^{-1}}$ \citep{Hobbs2005,FaucherGiguere2006,Igoshev2020,Disberg2025}, 
presumably imparted due to mass loss during the supernova explosion \citep{Blaauw1961,Hills1983,Kalogera1996,Kalogera1998,Kalogera2000}.
The subset of NSs that remain in binaries after core collapse, while a minority of the population, provides direct constraints on the physics of massive-star binary evolution and NS formation.

To date, several classes of NSs in binaries have been discovered in the Galaxy, including NSs with companions that are substellar objects, low-mass main-sequence (MS) stars ($\lesssim2~{\rm M_\odot}$), high-mass MS stars ($\gtrsim8~{\rm M_\odot}$), white dwarfs (WDs), and other NSs. 
NS--MS binaries with high-mass MS companions are often X-ray-selected and identified as (a) detached binaries or (b) actively accreting binaries.
Young pulsars with confirmed stellar companions have almost exclusively massive OB-star companions \citep{Kaspi1996,Lyne1994}, and detached X-ray-selected NS--MS binaries are classified as high-mass X-ray binaries, commonly with Be-star or supergiant donors \citep{Reig2011,Chaty2011,Fortin2023,Fortin2024,Liu2006,Liu2007}.

NS--MS binaries with low-mass MS companions, on the other hand, are commonly identified in various classes of close accreting binaries but have historically rarely been found in wide ($P_{\rm orb} \gtrsim 50$~days) detached binaries. In low-mass X-ray binaries (LMXBs), the NS accretes from a Roche-lobe-filling MS donor and exhibits periodic outbursts, allowing them to be identified in X-ray surveys. In spider pulsars, a recycled NS irradiates and ablates a very low-mass companion, creating a `day' and `night' side on the MS star that may be identified through photometric variability \citep{Bhattacharya1991,Podsiadlowski2002,Tauris2012,Roberts2013,Tauris2017}.

Recently, {\it Gaia} Data Release 3 (DR3) unveiled an entirely new NS population: detached NS--MS binaries containing solar-type companions in au-scale orbits \citep{ElBadry2024b,ElBadry2024d,ElBadry2026}. Unlike other known NSs\footnote{ Some systems could instead be unresolved WD+WD inner binaries with total masses in the NS range; this scenario is difficult to produce through standard triple evolution \citep[e.g.,][]{Shariat2023,Shariat2025c} but is not strictly excluded by current data (see Section~\ref{subsec:gaia_ns_sample} and \citealt{ElBadry2024b,ElBadry2026}).} 
with low-mass companions, these Gaia NS binaries show no evidence of past accretion, present accretion, or present pulsar activity \citep{ElBadry2024b,ElBadry2024d,Gao2026}. 
Their present-day orbits are wide and thus effectively identical to their post-supernova configurations, providing a probe of the core collapse process.

The formation history of Gaia NSs remains a puzzle. Given their current separation of $\sim1$~au, the NS progenitor is expected to have interacted with the $\sim1~{\rm M_\odot}$ companion when it was a red supergiant or super-AGB star.
At such extreme initial mass ratios ($q\lesssim0.1$), most models predict mass transfer to be dynamically unstable, leading to dramatic orbital shrinkage during a rapid common-envelope stage and, in many cases, a merger \citep{Paczynski1976,Webbink1984,Ivanova2013,Tauris2017}. 
How the Gaia NS progenitors avoided significant orbital contraction and remained wide upon emerging from mass transfer remains unclear. 
Even so, a wide post-mass-transfer binary must also survive NS formation, during which sudden mass loss and natal kicks can unbind the system or substantially reshape its orbit \citep{Blaauw1961,Hills1983,Kalogera1996,Kalogera1998}.

Dynamical formation in clusters could in principle bypass some of these difficulties, but this channel also appears inefficient for Gaia NSs. 
Open-cluster calculations find that most NSs are ejected by natal kicks before they can form Gaia NSs \citep{Tanikawa2024}, and disrupted dense-cluster models predict essentially no detectable au-scale NSs in Gaia DR3 \citep{SchiebelbeinZwack2026a}, inconsistent with the $27$ observed candidates.
Furthermore, several metal-poor Gaia NS candidates have extremely lithium-enhanced companions, suggesting that the luminous star interacted with the compact-object progenitor before or during the supernova \citep[e.g.,][]{ElBadry2024d}. This is difficult to reconcile with a late dynamical-exchange origin in clusters.

Despite the challenges posed by binary evolution, the Gaia NS binaries survived and emerged at wide separations with moderate eccentricities.  They therefore represent a novel outcome of extreme-mass-ratio binary evolution, distinct from the classical story of compact LMXBs, where NS--MS binaries previously underwent unstable common envelope evolution.
A promising pathway to disentangle the formation history of Gaia NSs is to constrain the intrinsic properties of their population: how common are Gaia NSs, and what is their distribution of orbital periods, eccentricity, MS masses, and NS masses.

However, the $N=27$ observed Gaia NS candidate sample is biased.  The {\it Gaia} non-single-star (NSS) catalog, from which Gaia NS candidates were selected, is shaped by the sensitivity of {\it Gaia} DR3, which reflects both intrinsic limitations of the data and cuts made by the non-single-star selection pipeline \citep{GaiaCollaboration2023,Holl2023,Halbwachs2023,ElBadry2024}.  The published Gaia NS sample also includes additional cuts imposed during candidate selection and spectroscopic follow-up \citep{ElBadry2024b,ElBadry2026}.  These selection effects, while biasing the data, can be quantified and leveraged to study the intrinsic properties of the underlying {\it Gaia} NS population. 

\citet{ElBadry2024} recently built a forward model of {\it Gaia} epoch astrometry and the astrometric-orbit pipeline, allowing the selection effects of {\it Gaia} DR3 binaries to be calculated and used to bias-correct the populations. The ability to accurately characterize the selection function of Gaia NSs makes them unique among other NS binaries. For example, the primary selection bias for NS-LMXBs, their X-ray recurrence timescale, is highly uncertain and varies widely from object to object \citep[e.g.,][]{Maccarone2022}.

In this paper, we carefully model the {\it Gaia} astrometric selection function to infer the intrinsic demographics of the Gaia NS population. Throughout the study, we assume the dark companions are NSs. If some candidates are instead massive WDs or unresolved WD+WD binaries, the inferred Gaia NS space densities and birth rates should be interpreted as upper limits. The remainder of the paper is organized as follows.
Section \ref{sec:methodology} outlines our methodology, including the selection function characterization and hierarchical Bayesian inference model. 
Section~\ref{sec:demographics} shows the best-fit results, including the inferred distribution of orbital period, eccentricity, visible-star mass, NS mass, and intrinsic space density of Gaia NS binaries. 
We also forecast the yields of Gaia NSs in {\it Gaia} DR4.
Section~\ref{sec:demographics_masses} discusses the mass distribution of Gaia NSs in the context of other NS binaries. 
Section~\ref{sec:ns_population_comparison} compares Gaia NSs to other Galactic NS populations, including isolated NSs, X-ray binaries, recycled pulsars, double neutron stars, and compact-object mergers.
Section~\ref{sec:conclusions} summarizes the main conclusions. Additional details are presented in Appendix \ref{app:corner_plot}, \ref{app:population_diagnostics}, \ref{app:high_mns}, and \ref{subsec:population_rate_estimates}.

\section{Methodology}\label{sec:methodology}

\subsection{The Gaia NS Sample}\label{subsec:gaia_ns_sample}

The observational sample studied in this paper is the 27 Gaia NS candidates identified by \citet{ElBadry2024b,ElBadry2026}.  These detached NS--MS candidates were first selected from the {\it Gaia} non-single-star (NSS) catalog using the astrometric mass-ratio function (AMRF), which combines the astrometric orbit, parallax, period, and visible-star mass to place a lower limit on the companion mass as a function of its light contribution \citep{Shahaf2019,Shahaf2024}.  In the AMRF diagram, ordinary single-MS binaries follow a main-sequence companion locus, while dark massive companions lie above this locus because they induce a large astrometric orbit without contributing much optical light.  The Gaia NS candidates occupy this dark-companion region, with inferred companion masses in the NS range ($\gtrsim1.25~{\rm M_\odot}$). From this initial selection, \citet{ElBadry2024b} and \citet{ElBadry2026} obtained RV follow-up to validate the astrometric solution and derive joint {\it Gaia}+RV orbital solutions. Several candidates were also identified as compact object candidates in previous {\it Gaia} DR3 samples \citep[e.g.,][]{Andrews2022b}.

The published Gaia NS sample from \citet{ElBadry2024b} and \citet{ElBadry2026} retained only sources with (a) $G<15$ to allow for precise RV follow-up, (b) a joint-fit companion mass $M_{\rm NS}>1.25~{\rm M_\odot}$, (c) RV coverage over at least half an orbit, and (d) published quality cuts that remove spurious astrometric solutions. They measure the orbital period ($P_{\rm orb}$), eccentricity ($e$), luminous-star mass ($M_1$), and NS candidate mass ($M_{\rm NS}$) for every Gaia NS candidate using their joint-fitting results.
The final sample spans $P_{\rm orb}=189$ -- $1405~{\rm d}$, $M_1=0.64$ -- $1.28~{\rm M_\odot}$, $M_{\rm NS}=1.28$ -- $1.88~{\rm M_\odot}$, and $e=0.12$ -- $0.79$, with typical uncertainties of $\approx5\%$ on $M_1$, $\approx3\%$ on $M_{\rm NS}$, $<0.3\%$ on periods, and $\approx1\%$ on eccentricities.

As discussed in the discovery papers \citep{ElBadry2024b, ElBadry2024c, ElBadry2026}, these binaries are NS {\it candidates} because the nature of the dark companion is uncertain. Fundamentally, the selection process constrains them to be dark companions (relative to the luminous star at optical wavelengths) with masses larger than $1.25~{\rm M_\odot}$. Plausible astrophysical sources satisfying these constraints are NSs, ultramassive WDs, or compact WD+WD binaries hidden in a hierarchical triple \citep{Shahaf2019,Shahaf2024,Hallakoun2024, Hallakoun2026,Garbutt2024}.  Compact double-WD binaries are generally difficult to explain on evolutionary grounds because the inner binary would need a near-Chandrasekhar total mass and to remain dynamically stable while the inner binary progenitors lost mass and expanded the outer orbit \citep{ElBadry2024b,ElBadry2024d,Toonen2016,Toonen2020,Shariat2023,Shariat2025c,Shariat2026_FRBs}.  

A single ultramassive WD is only a possibility for the $11$ objects below the Chandrasekhar limit \citep[$\approx1.37~{\rm M_\odot}$;][]{Althaus2022}, and these stars would need to have been drawn from the extreme high-mass tail of the WD mass distribution\footnote{Note the Chandrasekhar limit could be slightly larger ($\gtrsim1.4~{\rm M_\odot}$) for significantly rotating WDs \citep{Yoon2004}.}. $70\%$ 
($19/27$) of the candidates, including Gaia NS1 with $M_{\rm NS}=1.88\pm0.04~{\rm M_\odot}$ \citep{ElBadry2024d, ElBadry2026}, lie above $1.35~{\rm M_\odot}$, making them likely single NSs \citep{ElBadry2024b}. To test the impact of possible contamination from WDs, we repeat our analysis using only this high-mass subset of $M_{\rm NS}>1.35~{\rm M_\odot}$ (Appendix~\ref{app:high_mns}), finding that it does not change our qualitative conclusions. 
The eccentricities of Gaia NSs provide an additional clue: the candidates are more eccentric than ordinary WD--MS binaries at similar periods \citep{Shahaf2023,ElBadry2024b}.  We therefore adopt the simplest assumption that all systems in the sample are detached NS--MS binaries\footnote{If a fraction of the sample is composed of non-NS contaminants, then the inferred space densities and birth rates reported in this work should be interpreted as upper limits and can be scaled appropriately (e.g., $50\%$ contamination would lower the inferred rates by a factor of $2$).}.

\subsection{The {\it Gaia} selection function}\label{subsec:selection_function}

We quantify the {\it Gaia} DR3 selection function with {\tt gaiamock}\footnote{\url{https://github.com/kareemelbadry/gaiamock}}, a forward model of {\it Gaia} epoch astrometry and the astrometric-orbit pipeline \citep[see][for a detailed description]{ElBadry2024}. In short, {\tt gaiamock} implements the {\it Gaia} scanning law to generate mock epoch astrometry with an empirical noise model \citep{Holl2023}. It then replicates the astrometric selection pipeline used in {\it Gaia} DR3 to construct the published binary catalog \citep{Halbwachs2023}, where the mock astrometry is iteratively fit with increasingly complex orbital models until there is a sufficiently good fit, after which final quality cuts are made to remove potentially poor fits.
Previous works used {\tt gaiamock} to forward model the population of Gaia black holes (BHs) and Gaia WD--MS binaries \citep[][]{Nagarajan2025,Yamaguchi2025, Yamaguchi2026inprep}. Here, we apply similar methods to model the selection function for the Gaia NS sample.

To model the selection function of Gaia NSs, we execute a large injection--recovery campaign to estimate the probability that a binary with given physical parameters would be recovered by {\it Gaia} and pass the subsequent Gaia NS candidate cuts. The assumed distributions of the injected population are not important so long as they provide adequate coverage over the parameter space spanned by the observed sample \citep[e.g.,][]{Essick2025}. Our method is thus distinct from the classical forward model approach.

We start by injecting $140{,}000$ synthetic detached NS--MS binaries within $2$~kpc of the Sun.  The parent stars are drawn from a \texttt{Galaxia} Milky Way model \citep{Sharma2011}, which assigns their sky position ({\tt ra}, {\tt dec}), proper motion ({\tt pmra}, {\tt pmdec}), and parallax ($\varpi$). The apparent G-band magnitudes are derived for each injected source using MIST synthetic photometry, assuming no light contribution from the dark NS secondary. Line-of-sight extinction estimates are obtained using the \texttt{combined19} dust map \citep{Drimmel2003, Marshall2006, Green2019} from the \texttt{mwdust} package \citep{Bovy2016}. 
The metallicity ([Fe/H]) of each star is sampled from the skew-normal distribution of \citet{Zhang2023}, derived using the metallicity of stars in the solar neighborhood. The age is chosen uniformly between $0$ and $12$~Gyr to reflect a roughly constant star formation history during this period. The maximum age of a binary is chosen such that the primary is on the MS at present, given its $M_1$ and [Fe/H], with an upper bound of $12$~Gyr. The MS lifetimes for the companion stars are evaluated via interpolation of MIST evolutionary tracks \citep{Morton2015, Choi2016}. 

The injected population is split into three groups that together cover the entire observed sample. In all three groups, we generate injected binaries with uniformly sampled dark-companion masses over $1.1<M_2/{\rm M_\odot}<2.2$ and uniform eccentricity over $0<e<1$. The groups differ in orbital period and luminous star masses. 
The first set is $100{,}000$ injections over $100<P_{\rm orb}/{\rm d}<1300$ and $0.7<M_1/{\rm M_\odot}<1.4$. 
These original $10^5$ injections contained the full parameter space of the original Gaia NS sample \citep{ElBadry2024b}, but missed the longest-period and lowest-mass visible companions recently reported in the updated Gaia NS catalog \citep{ElBadry2026}. 
We therefore added $40{,}000$ additional injections to sufficiently cover the full parameter range.
$20,000$ injections span longer periods $1300<P_{\rm orb}/{\rm d}<1600$ and $0.55<M_1/{\rm M_\odot}<1.4$, and another $20{,}000$ span low-$M_1$ over $100<P_{\rm orb}/{\rm d}<1300$ and $0.55<M_1/{\rm M_\odot}<0.7$.
The orbital period of each binary is sampled from ${\rm d}N/{\rm d}P_{\rm orb}\propto P_{\rm orb}^{-0.2}$. 

As a reminder, the hierarchical Bayesian model (described in Section \ref{subsec:hbi_model}) is insensitive to the exact choice of priors used for the injected population, so long as they sufficiently sample the range of parameters spanned by the observed Gaia NS sample \citep[e.g.,][]{Essick2025,Yamaguchi2026inprep}. Our large injected population ($N=140{,}000$) and choice of broad priors for the population fits satisfy these prerequisites (e.g., Table \ref{tab:population_priors}).

\vspace{-0.8cm}
\begin{deluxetable*}{lcl}
\tablecaption{Cumulative cuts used for the selection function calculation; $N$ is the number of synthetic binaries remaining after each step.\label{tab:selection_cuts}}
\tablewidth{\textwidth}
\tablehead{
\colhead{Step} &
\colhead{Filter} &
\colhead{$N$}
}
\startdata
Initial Injection & -- & $140{,}000$ \\
Prefilter & $G_{\rm tot}<17.5$, $a_0/\sigma_{\rm AL}>1$ & $72{,}615$ \\
Recovered orbit & orbital solution from {\tt gaiamock} & $33{,}834$ \\
Parallax precision & $\varpi/\sigma_\varpi>20000/P_{\rm orb}$ & $15{,}745$ \\
Eccentricity precision & $\sigma_e < 0.079\ln P_{\rm orb}-0.244$ & $13{,}183$ \\
Photocenter significance & $a_0/\sigma_{a_0}>158/P_{\rm orb}^{1/2}$ & $9{,}660$ \\
Thiele-Innes precision & $\sigma_{\rm TI}^2\le36$ & $8{,}898$ \\
RV follow-up & $M_1\le1.35~{\rm M_\odot}$, $G<15$ & $6{,}756$ \\
NS candidate & AMRF boundary and $M_{{\rm NS},{\rm ast}}\ge1.25~{\rm M_\odot}$ & $5{,}727$ \\
\enddata
\end{deluxetable*}

Each injection is passed through two filters: (1) the {\tt gaiamock} astrometric cascade, which determines whether the binary would receive a non-single-star solution in {\it Gaia} DR3 \citep{ElBadry2024}, and (2) the extra NS candidate cuts imposed during catalog construction \citep{ElBadry2024b,ElBadry2026}. All cuts are summarized in Table~\ref{tab:selection_cuts}, showing that among the $140{,}000$ injected systems, $9{,}660$ pass the {\it Gaia} DR3 astrometric-quality cuts and $5{,}727$ pass all subsequent cuts for Gaia NS selection. 

Beyond the {\it Gaia} astrometric selection function, \citet{ElBadry2024b} impose additional cuts before selecting NS candidates, mainly to aid with RV follow-up. These include well-measured Thiele-Innes parameters (implicit from the AMRF selection), a bright low-mass visible star ($G<15$~mag), 
location above the single-main-sequence AMRF boundary of \citet{Shahaf2019,Shahaf2024}, and an astrometrically implied companion mass above $1.25~{\rm M_\odot}$.

The Thiele-Innes statistic measures the fractional uncertainty in the four Thiele-Innes constants of the {\it Gaia} astrometric orbit,
\begin{equation}
    \sigma_{\rm TI}^2=\sum_{K\in\{A,B,F,G\}}\left(\frac{\sigma_K}{K}\right)^2 .
\end{equation}
We require $\sigma_{\rm TI}^2\le36$ following \citet{Shahaf2024}, which leaves $8{,}898$ injections.  The remaining candidate-window cuts require $M_1\le1.35~{\rm M_\odot}$, $P_{\rm orb}\le1600~{\rm d}$, and $G<15$, matching the published sample window and leaving $6{,}756$ injections \citep{ElBadry2024b}.

The final selected sample contains $5{,}727$ injections that satisfy $M_{{\rm NS},{\rm ast}}\ge1.25~{\rm M_\odot}$ and fall above the AMRF boundary.
These recovered systems define our best approximation to the Gaia NS sample selection and are used to model the selection function.
The likelihood of our Bayesian model (Section \ref{subsec:hbi_model}) considers the complete selection function: that a binary needs to both get a DR3 solution and pass the subsequent follow-up selection and quality cuts. 

We note that the observed sample was originally assembled from different compact-object parent samples, including candidates identified in earlier {\it Gaia} compact object searches \citep[e.g.,][]{Andrews2022b}, whose selection steps are not straightforward to reproduce in an injection campaign. In addition, the published $M_{\rm NS}>1.25~{\rm M_\odot}$ threshold was applied to the final joint {\it Gaia}+RV companion masses, whereas our synthetic systems only consider astrometrically inferred masses. Thus, the selection function used here should be understood as a best approximation to the published Gaia NS sample and captures the dominant DR3 selection effects.  Crucially, the selection function model presented here can be readily applied to {\it Gaia} DR4 samples as they are released.

\subsubsection{Summary}\label{subsubsec:selection_function_summary}

\begin{figure*}
    \centering
    \includegraphics[width=0.99\linewidth]{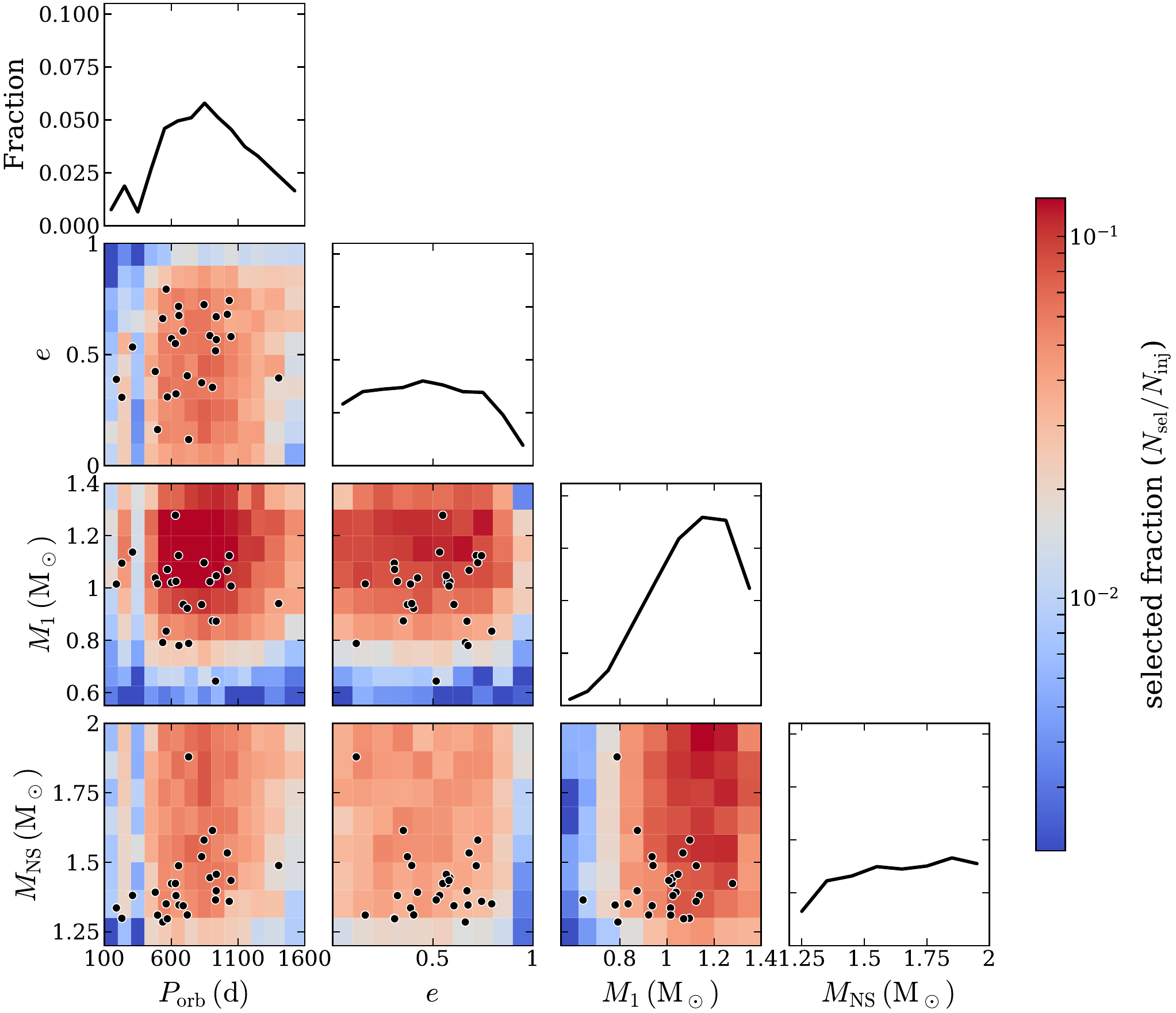}
    \caption{Selection function of Gaia NS systems in {\it Gaia} DR3.  Diagonal panels show the 1D selected fraction among injection systems. Lower panels show the two-dimensional selected fraction, with the observed Gaia NS candidates overplotted (black circles). Gaia NSs are largely found in high-probability regions of parameter space, showing that some observed properties are a result of the selection function.
    }
    \label{fig:selection_function}
\end{figure*}

Figure \ref{fig:selection_function} shows the selection function of our Gaia NS search as a function of orbital period, eccentricity, visible-star mass, and NS mass. The plots on the diagonal show the sensitivity as a function of each parameter individually, and the off-diagonal plots show the sensitivity in the two-dimensional parameter spaces. The colors show the fraction of injected systems that were recovered and selected as Gaia NS candidates. The observed Gaia NS candidates are shown in black.
This figure highlights the key features of the {\it Gaia} DR3 selection function for NS-harboring binaries. 

{\it Gaia} DR3 is most sensitive to astrometric binaries with periods that are well sampled over its $\sim10^3$~day observing baseline (Figure \ref{fig:selection_function}). There is reduced sensitivity at short periods ($\lesssim300$~days) because the angular on-sky orbit is small and at long periods ($\gtrsim900$~days) because the orbit becomes poorly covered as the period approaches or exceeds the baseline \citep{GaiaCollaboration2023,Halbwachs2023,Holl2023}.
A discrete drop in sensitivity is observed near $P_{\rm orb}\approx365$~days, which arises from the degeneracy with parallactic motion. 

The astrometric sensitivity is constant from low to moderate eccentricities (Figure \ref{fig:selection_function}), but decreases sharply at high eccentricities ($e\gtrsim0.7$). At high eccentricity, most of the photocenter motion is concentrated into a small fraction of the orbit near periastron, making the astrometric solution more sensitive to the scanning, which depends on location.
Sensitivity increases toward larger visible-star masses ($M_1$) because these systems are typically brighter and thus have smaller astrometric uncertainties. The selection function also mildly favors larger NS masses ($M_{\rm NS}$). This trend reflects both the larger astrometric signal induced by more massive companions and the AMRF selection cuts. Namely, systems with larger inferred $M_{\rm NS}$ induce more astrometric motion at fixed orbital period and $M_1$, and they also rule out a larger range of possible luminous companions, satisfying the AMRF cut.

Comparing the observed Gaia NS candidates in Figure \ref{fig:selection_function} to the detection probability reveals intrinsic features of the population. Detections in low-sensitivity regions carry more weight in an inference model because they suggest that many systems likely intrinsically reside in that region of parameter space, given some were observed.
For example, there are three Gaia NS candidates with $P_{\rm orb}<350$~days, despite the substantially reduced sensitivity at short periods, suggesting that compact systems may be intrinsically more abundant than implied by the raw detections alone (Figure \ref{fig:selection_function}). Conversely, only $5$ candidates have $M_{\rm NS}>1.5~{\rm M_\odot}$, even though the selection function modestly favors higher NS masses through both larger astrometric signals and the AMRF cuts. This may already suggest that high-mass NS candidates are intrinsically rare in Gaia NS binaries.
Only $1$ system has a luminous star with $M_1>1.2~{\rm M_\odot}$, despite high sensitivity in that region. This likely reflects the shorter main-sequence lifetimes of higher-mass companions, which make them less common among older Gaia NS binaries.
Overall, Figure~\ref{fig:selection_function} provides an instructive visualization of the main selection effects that bias the observed sample. We quantify these trends in the hierarchical Bayesian model described below. 

\subsection{Hierarchical Bayesian Model}\label{subsec:hbi_model}

We model the data as an inhomogeneous Poisson point process in $x=(P_{\rm orb},M_1,M_{\rm NS},e)$, following standard selection-corrected population inference \citep{Loredo2004,Mandel2019,ForemanMackey2014,Wysocki2019,Fishbach2018,Essick2024,Essick2025}.  This implementation adapts the hierarchical Bayesian model framework of \citet{Yamaguchi2026inprep} to the Gaia NS candidate sample, which is based on the works of \citet{Essick2024} and \citet{Essick2025}. 
We refer readers to these works for the full model description and derivations, and summarize here the background and key aspects needed for this analysis, including modifications specific to the Gaia NS sample.

Our goal is to infer properties of the {\it intrinsic} Gaia NS population.
The effective DR3 selection fraction is
\begin{equation}
    \beta(\Lambda)=P(D\mid\Lambda)=\int p(D\mid x)~p(x\mid\Lambda)~dx.\label{eq:beta_eff}
\end{equation}
Here, $x$ denotes the parameters of an individual binary (orbital period, eccentricity, masses). A population model $\Lambda$ specifies the intrinsic distribution of these parameters, $p(x\mid\Lambda)$. For a binary with parameters $x$, $p(D\mid x)$ is the probability that it is detected in our modeled Gaia NS sample, including both receiving an astrometric orbital solution in DR3 and passing the additional Gaia NS candidate cuts described above.

We estimate this integral using the injected population and importance reweighting:
\begin{equation}
    \widehat{\beta}(\Lambda)=\frac{1}{N_{\rm inj}}
      \sum_{j=1}^{N_{\rm inj}} I_j\frac{p(x_j\mid\Lambda)}{p_{\rm inj}(x_j)}.
    \label{eq:beta_reweighting}
\end{equation}
Here $x_j$ denotes the parameters of the $j$th injected binary, $p_{\rm inj}(x_j)$ is the probability of selecting that binary from the injected distribution, and $p(x_j\mid\Lambda)$ is the probability density assigned to the same binary by the population model being tested. The indicator $I_j=1$ if the injected binary is selected as a Gaia NS candidate and $0$ otherwise. The ratio $p(x_j\mid\Lambda)/p_{\rm inj}(x_j)$ therefore reweights each injection from the artificial injection distribution to the intrinsic population model. This lets us evaluate the selection fraction for many different population models using a single injection campaign, provided the injections cover the region of parameter space where $p(x\mid\Lambda)$ is appreciable.

The corresponding log-likelihood is
\begin{equation}
\begin{split}
    \log \mathcal{L}
    =&
    -K_{\rm 2kpc}\beta(\Lambda)
    + N_{\rm det}\log K_{\rm 2kpc} \\
    &+
    \sum_{i=1}^{N_{\rm det}}
    \log\left[
    \int p(d_i\mid x)\,p(x\mid\Lambda)\,dx
    \right],
\end{split}
\label{eq:hbi_log_likelihood}
\end{equation}
where $N_{\rm det}=27$, $K_{\rm 2kpc} \equiv R$ is the intrinsic number of Gaia NS-like systems within 2 kpc, and $\beta(\Lambda)$ is the selection fraction for population parameters $\Lambda$.  The quantity $K_{\rm 2kpc}\beta(\Lambda)$ is the expected number of selected DR3 systems.  The first two terms are the Poisson likelihood for detecting $N_{\rm det}$ systems, while the sum fits where the detected systems fall in $(P_{\rm orb},M_1,M_{\rm NS},e)$.
The model is therefore used not only to constrain the distributions of periods, masses, and eccentricities, but also the intrinsic number of candidates within 2 kpc.  

Measurement uncertainties are incorporated into the likelihood through Monte Carlo integration over the observed systems.  For each candidate, we draw 96 samples from independent Gaussian uncertainties in $P_{\rm orb}$, $M_1$, $M_{\rm NS}$, and $e$.  These samples are drawn once and held fixed throughout the MCMC.  We treat the four uncertainties as independent because the published joint astrometry+RV fits provide small ($\sim1$--$5\%$) uncertainties compared to the population widths and because the dominant uncertainty in this paper is the selection correction. 

\vspace{-0.8cm}
\begin{deluxetable*}{llll}
\tablecaption{Population model and priors.\label{tab:population_priors}}
\tablehead{
\colhead{Quantity} & 
\colhead{Distribution} & 
\colhead{Range} &
\colhead{Prior}
}
\startdata
$P_{\rm orb}$ 
& Power law, $p(P_{\rm orb})\propto P_{\rm orb}^{\alpha_P}$ 
& $100$ -- $1600~{\rm d}$ 
& $\alpha_P\sim U(-4,4)$ \\
$M_1$ 
& Truncated normal, $(\mu_{M_1},\sigma_{M_1})$ 
& $0.55$ -- $1.4~{\rm M_\odot}$ 
& $\mu_{M_1}/{\rm M_\odot}\sim U(0.55,1.4)$; $\ln\sigma_{M_1}\sim U(-4,-0.05)$ \\
$M_{\rm NS}$ 
& Truncated normal, $(\mu_{M_{\rm NS}},\sigma_{M_{\rm NS}})$ 
& $1.1$ -- $2.2~{\rm M_\odot}$ 
& $\mu_{M_{\rm NS}}/{\rm M_\odot}\sim U(1.1,2.2)$; $\ln\sigma_{M_{\rm NS}}\sim U(-4,-0.05)$ \\
$e$ 
& Beta distribution, ${\rm Beta}(\mu_e\kappa,[1-\mu_e]\kappa)$ 
& $0$ -- $1$ 
& $\mathrm{logit}\,\mu_e\sim U(-8,8)$; $\ln(\kappa-2)\sim U(-2,6)$ \\
$R$ 
& Poisson process 
& -- 
& $\ln R\sim U(-50,50)$ \\
\enddata
\end{deluxetable*}

\subsection{Population Parameterization}\label{subsec:population_model}

Our fiducial model\footnote{Alternative parameterizations are tested in Appendix~\ref{app:population_diagnostics}; we find that our results are robust to such choices.} assumes that the joint probability distribution is separable, or equivalently, that the individual distributions are independent:
\begin{align}
    \begin{split}
    p(x|\Lambda) &= p(P_{\rm orb}, M_1, M_2, e|\Lambda)   \\ &=p(P_{\rm orb}|\Lambda)p(M_1|\Lambda)p(M_2|\Lambda)p(e|\Lambda)
    \end{split}
\end{align}
We adopt this simple fiducial model for the 27-object sample.
The posterior predictive checks in Section~\ref{subsec:intrinsic_demographics} show that it reproduces the observed distributions reasonably well, suggesting that the assumption is adequate at the current sample size.
Future {\it Gaia} data releases with larger NS samples warrant further investigation of this assumption.

We assume the period distribution is a power law,
\begin{equation}
    p(P_{\rm orb})\propto P_{\rm orb}^{\alpha_P},
    \qquad 100<P_{\rm orb}/{\rm d}<1600 .
\end{equation}
The visible-star and NS companion masses are assumed to take the form of truncated normal distributions over their entire range $0.55<M_1/{\rm M_\odot}<1.4$ and $1.1<M_{\rm NS}/{\rm M_\odot}<2.2$.  Eccentricity is modeled as a Beta distribution parameterized by mean $\mu_e$ and concentration $\kappa$,
\begin{equation}
    p(e)=\mathrm{Beta}\left(e\mid \mu_e\kappa,(1-\mu_e)\kappa\right),
    \qquad 0<e<1 .
\end{equation}
All distributions $p(x)$ are normalized such that they integrate to unity.
The sampled parameter vector is
\begin{equation}
\begin{split}
    \theta=(&\alpha_P,\mu_{M_1},\log\sigma_{M_1},
    \mu_{M_{\rm NS}},\log\sigma_{M_{\rm NS}},\\
    &\mathrm{logit}~\mu_e,\log(\kappa-2),\log R).
\end{split}
\end{equation}
We sample $\mathrm{logit}\,\mu_e=\log[\mu_e/(1-\mu_e)]$ instead of $\mu_e$ directly so that the eccentricity mean is restricted to $0<\mu_e<1$ while the sampled parameter can vary over the full real line.  Similarly, sampling $\log(\kappa-2)$ enforces $\kappa>2$, which avoids extremely singular eccentricity distributions at the boundaries. The population model and priors are summarized in Table \ref{tab:population_priors}.

\subsection{MCMC Inference and Posterior Predictive Checks}\label{subsec:mcmc_ppc}

We sample the likelihood (Equation~\ref{eq:hbi_log_likelihood}) using the \texttt{emcee} package \citep{ForemanMackey2013}.  We use $72$ walkers, $800$ burn-in steps, and $18{,}000$ production steps.  We inspect trace plots, posterior corner plots, and the stability of the selected-count distribution before using the chains.

Posterior predictive checks (PPCs) are performed as follows.  For each posterior draw, we reweigh injections passing the final NS candidate selection by $p(x\mid\Lambda)/p_{\rm inj}(x)$ and draw mock selected catalogs with the same size as the observed sample.  We then compare the mock and observed distributions in $P_{\rm orb}$, $M_1$, $M_{\rm NS}$, and $e$.  This check asks whether the inferred intrinsic population, after the same selection function is applied, reproduces the actual selected sample. 

\begin{figure}
    \centering
    \includegraphics[width=0.95\linewidth]{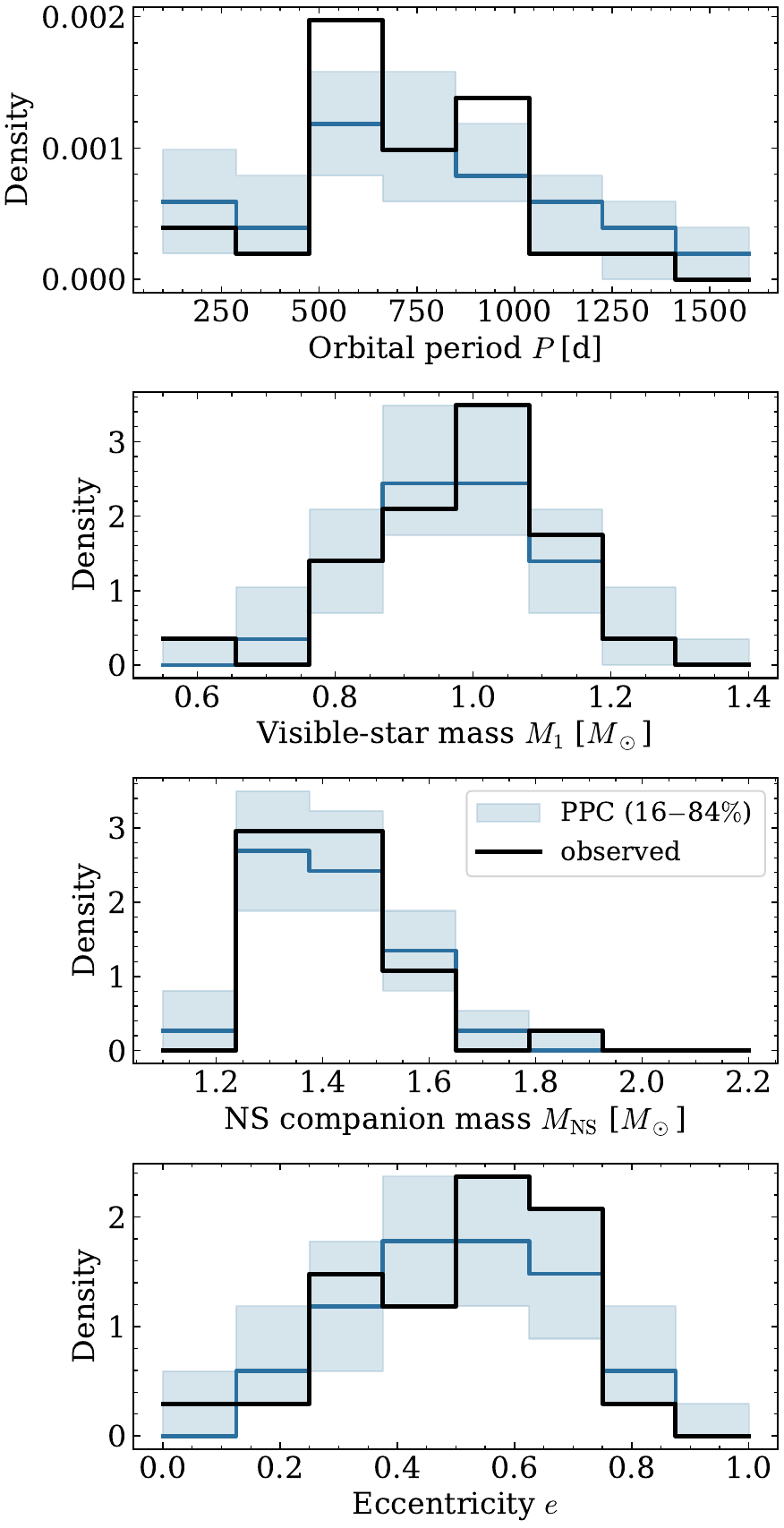}
    \caption{ Posterior predictive checks (PPC) from our hierarchical Bayesian model. Blue histograms show the predicted Gaia NS sample from the model's posterior ($16$ -- $84\%$ ranges shaded), compared to the observed Gaia NS population (black). The resulting distributions of orbital period, visible-star mass, NS mass, and eccentricity are shown and broadly agree well with the data. The consistency suggests that the choice of parameterization reasonably approximates the Gaia NS population in DR3.}
    \label{fig:PPC}
\end{figure}

As illustrated in Figure \ref{fig:PPC}, the model reasonably matches the observed population.
In this figure, we sample periods, visible star masses, NS masses, and eccentricities from the best-fit posterior population for many realizations, finding that the resulting distributions are consistent with the observed Gaia NS population within $1\sigma$ uncertainties. The consistency further supports our choice of parameterization for each distribution (Section \ref{subsec:population_model}), since they seemingly provide a reasonable fit to the current 27-object sample.
The same draws give the expected number of selected systems in Gaia DR3, 
$\mu = R\beta(\Lambda)$, which should be similar to the observed number of Gaia NSs ($N=27$), within uncertainty. We indeed find the model predicts $\mu=26.6^{+5.5}_{-4.9}$.

Our fiducial model assumes that parameters have independent intrinsic distributions, but the selection function can produce correlations among the detected parameters. We therefore compared the correlation in each of the six 2D parameter combinations with the correlations measured in PPC samples. Five are consistent with the model. The exception is $P_{\rm orb}$ and $M_{\rm NS}$: the observed sample has $\rho_{\rm S}=0.58$, stronger than predicted by the model. This correlation remains after propagating measurement uncertainties and repeating the calculation while removing each system in turn. It may indicate an intrinsic relationship not captured by our separable model, although the heterogeneous selection of the candidate sample could also contribute. A larger sample in Gaia DR4 will help solidify any correlations.

From the intrinsic number of Gaia NSs predicted within $d_{\max}=2~{\rm kpc}$, we calculate the local midplane density
\begin{equation}
    n_0=\frac{N_{2~{\rm kpc}}}{V_{\rm eff,local}},
\end{equation}
where the effective volume is evaluated relative to the density at the Solar position,
\begin{equation}\label{eq:v_eff_local}
\begin{aligned}
V_{\rm eff,local} ={}&\int_{d<d_{\max}}
\exp\left[-\frac{R(d,l,b)-R_0}{R_d}\right] \\
&\qquad\times\exp\left[-\frac{|z(d,b)|}{h_z}\right] dV .
\end{aligned}
\end{equation}
Here $R_d$ is the scale length of the thin disk, and  $R(d,l,b)$ is the Galactocentric cylindrical radius of a point at heliocentric distance $d$, Galactic longitude $l$, and latitude $b$.  We adopt $R_0=8.2~{\rm kpc}$, $R_d=2.6~{\rm kpc}$, and $h_z=300~{\rm pc}$ \citep{Bovy2016, McMillan2017}\footnote{Note this does not capture Gaia NSs belonging to the Galactic halo, which currently comprises a minority of the observed population ($3/27$).}.  Integrating over the local $d<2~{\rm kpc}$ volume gives $V_{\rm eff,local}\approx7.6~{\rm kpc^3}$. 

To convert this local density to a total Galactic number, we integrate the same density model over the thin disk,
\begin{equation}\label{eq:v_eff_galactic}
\begin{aligned}
V_{\rm eff,Gal} ={}&2\pi\int_0^{R_{\rm max}}
\int_{-\infty}^{\infty} R\, \\
&\quad\times\exp\left[-\frac{R-R_0}{R_d}-\frac{|z|}{h_z}\right] dR\,dz.
\end{aligned}
\end{equation}
For $R_{\rm max}=13~{\rm kpc}$, this gives $V_{\rm eff,Gal}\approx570~{\rm kpc^3}$, so $N_{\rm Gal}=n_0V_{\rm eff,Gal}$ under the adopted exponential-disk model.

\subsection{{\it Gaia} DR4 Forecast Inputs}\label{subsec:dr4_forecast_inputs}
We forecast the number of detached NS--MS binaries that could appear in {\it Gaia} DR4 by forward-modeling the DR3-inferred population through the DR4 selection process using {\tt gaiamock}. 
The mock observations generally follow the procedure outlined by \citet{Nagarajan2025} for {\it Gaia} BH forecasts, but with the intrinsic binary population supplied by our {\it Gaia} DR3 NS model.   
Each NS-hosting star is assigned an observed sky position, distance, proper motion, parallax, and apparent $G$ magnitude using the simulated galaxy {\tt m12i} from the FIRE-2 simulation suite \citep{Wetzel2016,Hopkins2018}. 3D Galactic positions are assigned to stars  based
on the {\tt ananke} framework of \citet{Sanderson2020}, which
generates mock catalogs of synthetic stars from cosmological
baryonic simulations \citep[see][for more details]{Chawla2022}. Gaia NSs are then randomly assigned to star particles. Extinction is evaluated with the \texttt{combined19} dust map in \texttt{mwdust}, which combines the maps of \citet{Drimmel2003}, \citet{Marshall2006}, and \citet{Green2019} .  The dark companion is assumed to contribute no optical light, so the simulated astrometric signal is the photocenter orbit of the luminous star.

For each binary in the DR4 realization, we draw the orbital orientations and phases at random: $\cos i$ is uniform, $\omega$ (argument of periastron), $\Omega$ (longitude of the ascending node), and the periastron time are uniform over their allowed ranges.  {\tt gaiamock} then uses the {\it Gaia} scanning law to generate epoch astrometry with the appropriate along-scan sampling and astrometric uncertainties, including the orbital displacement of the photocenter \citep{Lindegren2022,Holl2023,ElBadry2024}.  The mock epoch astrometry is passed through the {\it Gaia} astrometric model cascade, which tests increasingly complex solutions and returns whether the source receives a single-star, acceleration, or orbital solution, following the logic of the DR3 non-single-star catalog construction \citep{Halbwachs2023,ElBadry2024}.  The DR4 run allows for orbital periods up to  $P_{\rm orb}=10^4~{\rm d}$. For the forecasts presented here, we only consider $d<10~{\rm kpc}$ and $P_{\rm orb}<7000~{\rm d}$, where the simulated long-period orbit solutions are best characterized \citep{Nagarajan2025}.  

Our original {\it Gaia} DR3 sample was restricted to $d<2$~kpc, so we extend to larger distances for DR4 in a manner that preserves the inferred space density in DR3.  
We allow for two definitions of a `DR4 detection' following \citet{Nagarajan2025}, which we similarly categorize as (1) `same cuts as DR3' and (2) `less stringent.'
The first applies the same period-dependent orbit-quality cuts used for {\it Gaia} DR3 in Section~\ref{subsec:selection_function}.  This is the conservative choice, because the DR3 non-single-star catalog was filtered aggressively to suppress spurious orbital solutions.  In particular, \citet{Halbwachs2023} introduced a period-dependent parallax-significance cut,
\begin{equation}
    \frac{\varpi}{\sigma_\varpi} > \frac{20000\,{\rm d}}{P_{\rm orb}},
\end{equation}
to remove dubious astrometric orbital solutions concentrated near particular periods.  The second, `less stringent' definition requires only a valid {\tt gaiamock} orbital solution, $\varpi/\sigma_\varpi>5$, and $a_0/\sigma_{a_0}>5$, where $a_0$ is the photocenter semimajor axis.  This cut keeps systems with significant parallaxes and significant orbital photocenter motion, but does not impose the DR3 period-dependent rejection filter. 
Applying the looser cuts to real DR4 data will likely result in a significant number of false positives, where radial velocity follow-up is required to confirm their orbits.

The main extrapolation from DR3 to DR4 for the Gaia NS population is the period distribution beyond $\sim1000$~d.  For our DR4 forecast, we accommodate longer periods by extending the fitted single power law to $P_{\rm orb}=7000$~d from the original $P_{\rm orb}=1600$~d in the DR3 sample. To avoid blind extrapolation, we allow the long-tail period distribution to follow an independent power law with slope $\beta$:
\begin{equation}
    p(P_{\rm orb}) \propto
    \begin{cases}
    P_{\rm orb}^{\alpha_P}, & P_{\rm orb} \leq1600~{\rm d},\\
    P_{\rm break}^{\alpha_P-\beta}P_{\rm orb}^{\beta}, & P_{\rm orb}>1600~{\rm d},\label{eq:dr4_power_law}
    \end{cases}
\end{equation}
where $P_{\rm break}=1600$ d. The $P_{\rm break}^{\alpha_P-\beta}$ prefactor keeps the period distribution continuous at the break. For the low-period branch, ${\alpha_P}$ is sampled from the posterior inferred from DR3.

\section{Population Properties}\label{sec:demographics}

\subsection{Orbital periods, masses, and eccentricities}\label{subsec:intrinsic_demographics}

\begin{figure*}
    \centering
    \includegraphics[width=0.99\linewidth]{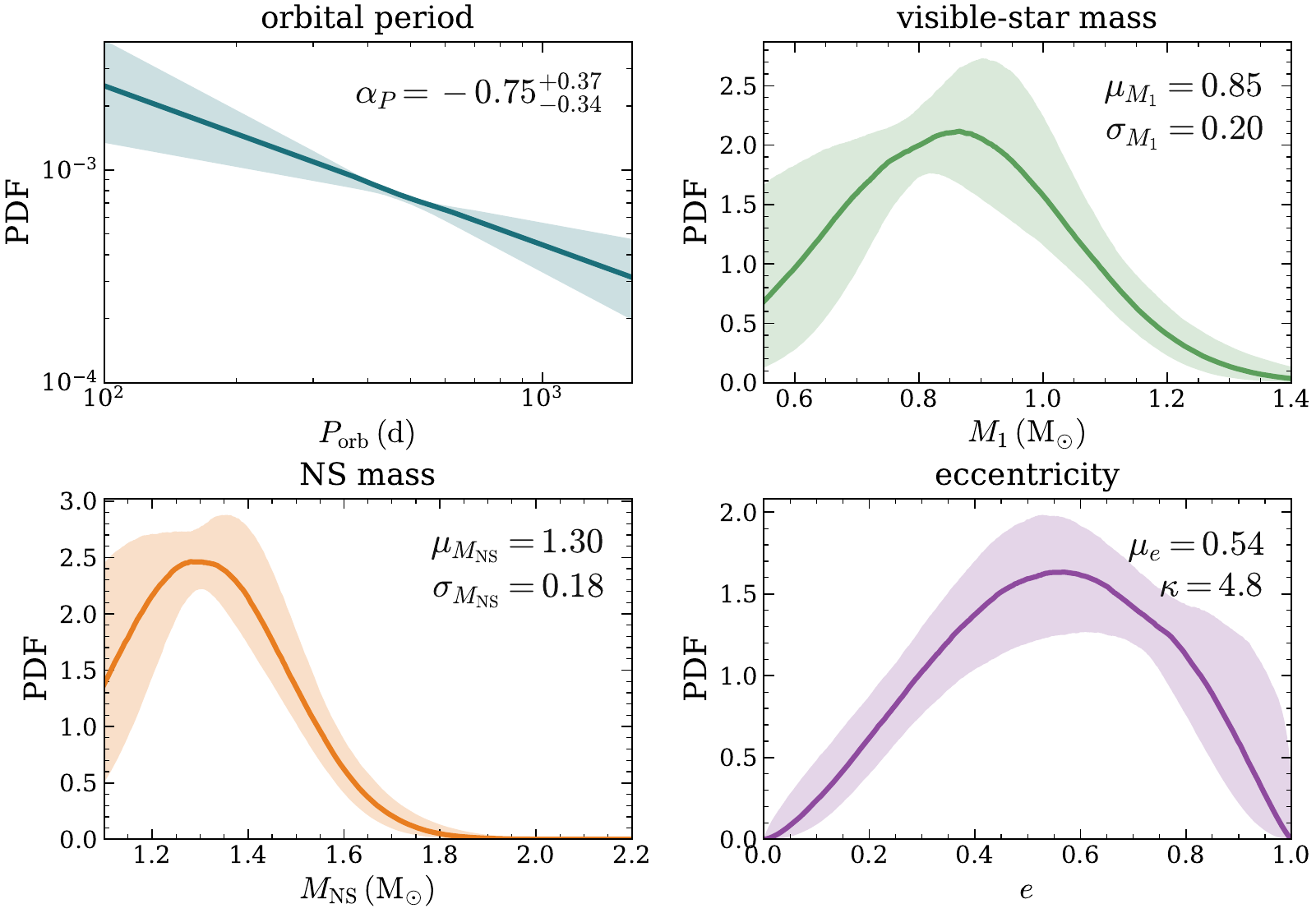}
    \caption{Intrinsic distributions inferred for the {\it Gaia} DR3 NS--MS candidate population.  We show posterior medians (solid line) and 16th--84th percentile intervals (shaded regions) from the hierarchical Bayesian model, displayed as probability density functions. In each panel, we also write the best-fit model parameters, which define the distributions. }
    \label{fig:best_fit_distributions}
\end{figure*}

Figure~\ref{fig:best_fit_distributions} shows the resulting intrinsic properties of the Gaia NS population. The posterior probability density functions are displayed with the best fit and $68\%$ ($1\sigma$) interval shown (see Appendix \ref{app:corner_plot} for the full corner plot).
The orbital period distribution of Gaia NSs is weighted toward short periods with best-fit power law slope $\alpha_P=-0.75^{+0.37}_{-0.34}$. A decreasing period distribution is favored by the data and mainly driven by the fact that $3$ observed Gaia NSs reside at $P_{\rm orb}<400$~d despite poor sensitivity there (Figure \ref{fig:selection_function}). The preference for a decreasing intrinsic period distribution is also robust to the specific choice of parameterizations (Appendix \ref{app:population_diagnostics}), albeit with relatively large uncertainties.  

The NS mass distribution peaks at $\mu_{M_{\rm NS}}=1.30^{+0.07}_{-0.11}~{\rm M_\odot}$ with spread $\sigma_{M_{\rm NS}}=0.18^{+0.05}_{-0.04}~{\rm M_\odot}$.  This peak is broadly similar to the peak mass scale found in other NS populations, including accreting NSs, recycled pulsars, and double NS binaries \citep{Valentim2011,Kiziltan2013,Antoniadis2016,Alsing2018,ElBadry2024b,You2025,SchiebelbeinZwack2026}, which generally place a peak in NS masses near $1.3~{\rm M_\odot}$ (see Section \ref{sec:demographics_masses} for further discussion).

Recall that the published Gaia NS sample is restricted to $M_{\rm NS}>1.25~{\rm M_\odot}$, although lower-mass NSs are known in other binaries \citep[e.g.,][]{Ferdman2014,Martinez2015}.
The inferred behavior near and below $1.25~{\rm M_\odot}$ should therefore be interpreted with caution, and does not mean that the {\it Gaia} data directly measure the mass distribution below the selection threshold. Hence the large $1\sigma$ scatter in this region.
Rather, within the mass range sampled by the observed candidates, the likelihood favors an intrinsic distribution that decreases toward higher $M_{\rm NS}$.
In the fiducial Gaussian model, this decreasing trend is represented by a mean near the lower edge of the observed mass window.
This result is driven by the fact that the Gaia selection function mildly favors higher-mass companions, yet the observed sample contains relatively few high-mass NS candidates.
Thus, after correcting for selection, the intrinsic Gaia NS mass distribution is pushed toward lower masses.
This preference persists when we restrict the analysis to the high-mass subset, $M_{\rm NS}>1.35~{\rm M_\odot}$, and when we use alternative mass parameterizations beyond a truncated Gaussian (Appendices~\ref{app:high_mns} and~\ref{app:population_diagnostics}). 

Notably, the Gaia NS sample provides a rather practical new population for measuring the NS mass distribution.  Most precise NS masses prior to {\it Gaia} were derived from pulsars in compact binaries or accreting X-ray binaries \citep{OzelFreire2016,Antoniadis2016}.  These measurements are generally precise but can be biased tracers of the birth mass due to prior accretion \citep[e.g.,][]{You2025}.
The NSs in wide Gaia NS binaries, on the other hand, are thought to have had no significant accretion onto the NS, making them a relatively pristine measure of the NS birth-mass distribution.

The visible-star mass distribution of the Gaia NS population is centered slightly below a solar mass, with $\mu_{M_1}=0.85^{+0.07}_{-0.14}~{\rm M_\odot}$ and $\sigma_{M_1}=0.19^{+0.09}_{-0.05}~{\rm M_\odot}$.  This preference for sub-solar-mass companions is likely partly due to astrophysical biases.  Lower-mass stars are more common in the field due to the IMF\footnote{although the mass-ratio distribution is poorly constrained at $q<0.1$ \citep[e.g.,][]{Moe2017}, so the abundance of $\sim1~{\rm M_\odot}$ companions around O- and B-type stars is unknown.} and remain on the main sequence for longer. At roughly solar metallicity, a $0.8~{\rm M_\odot}$ star has a main-sequence lifetime comparable to or longer than the age of the Universe, while a $1.2~{\rm M_\odot}$ star leaves the main sequence after only $\sim5$ -- $6$ Gyr \citep[e.g.,][]{Choi2016}.  Thus, lower-mass luminous companions are expected to be overrepresented among older Gaia NS binaries.  

The eccentricity distribution is broad and favors moderate eccentricities, with $\mu_e=0.54^{+0.06}_{-0.05}$ and $\kappa=4.8^{+2.7}_{-1.7}$.  The high-eccentricity tail ($e>0.7$) remains poorly constrained because {\it Gaia} has low sensitivity in this region (Figure~\ref{fig:selection_function}).  
Noticeably, the distribution has little support for circular or nearly circular orbits.  This is consistent with the NS interpretation, where mass loss and natal kicks during the supernova produce eccentricity.  In contrast, most WD--MS binaries at these periods have low eccentricities, as is nominally expected if they experienced tidal circularization when the WD progenitor was a giant. However, several post-interaction WD--MS populations have nonzero eccentricities, including those on au-scale orbits.  This could indicate mass transfer beginning on the asymptotic giant branch, unstable common-envelope evolution, or eccentricity pumping mechanisms during mass transfer \citep{Phinney1992,Lorimer2008,Mathieu2009,Jorissen2019,Escorza2019,Yamaguchi2024,Yamaguchi2025,vanSon2026}.

\subsection{Space density}\label{subsec:space_density}

\begin{figure}
    \centering
    \includegraphics[width=0.99\columnwidth]{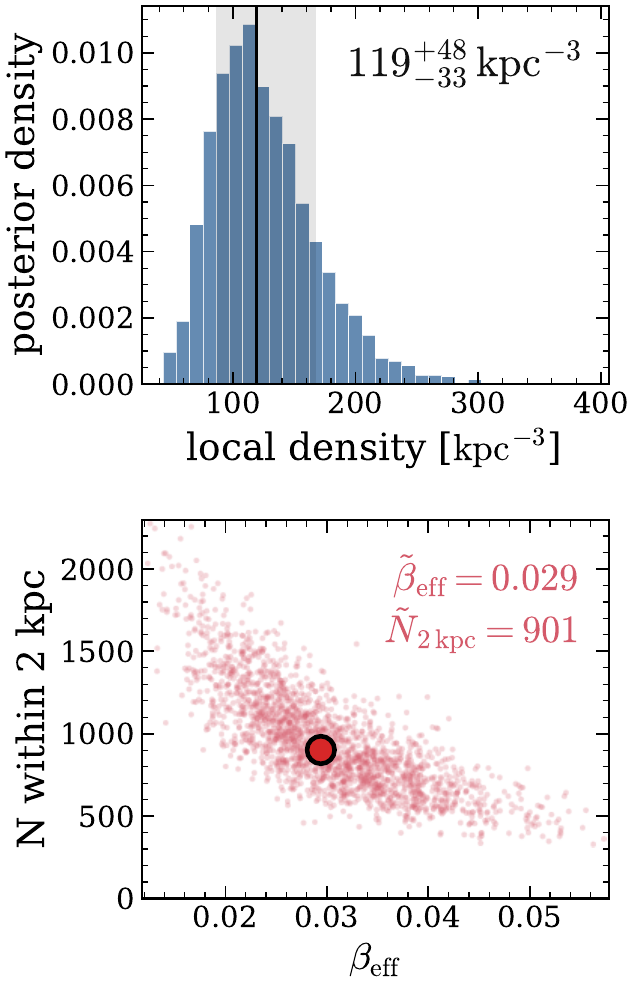}
    \caption{Intrinsic space density of Gaia NS binaries. The top panel shows the inferred intrinsic space density obtained by sampling the posteriors, with a best-fit value of $n=119~{\rm kpc^{-3}}$. The bottom panel shows the predicted total number of Gaia NS binaries within $2$~kpc ($N_{\rm 2~kpc}$) compared to the fraction detected in {\it Gaia} DR3 ($\beta_{\rm eff}$). $N_{\rm 2~kpc}$ is inversely correlated with the detection fraction $\beta_{\rm eff}$, and $N_{\rm 2~kpc}\times\beta_{\rm eff} \approx 27$, consistent with the DR3 sample.}
    \label{fig:rate_normalization}
\end{figure}

A result of modeling the selection function and performing hierarchical Bayesian inference is an estimate of the true number of objects in the parent sample.
Figure~\ref{fig:rate_normalization} shows the total number of Gaia NSs within $2$~kpc, as predicted by our population model.
As shown in the bottom panel, our model predicts $N_{2~{\rm kpc}}=900^{+366}_{-251}$ Gaia NS binaries within 2 kpc in the range of periods, eccentricities, and masses spanned by the observed sample. The total estimated number is sensitive to the selection function model, and thus shows an inverse correlation with the total selection fraction detected in {\it Gaia} DR3, $\beta_{\rm eff}$ (Equation \ref{eq:beta_eff} and bottom panel of Figure~\ref{fig:rate_normalization}). Sampling from our model posteriors and applying the DR3 selection function predicts $\mu=26.6^{+5.5}_{-4.9}$ Gaia NSs observed in DR3, consistent with the true value of $N=27$ within Poisson uncertainties, and providing an encouraging posterior predictive check. 
The fact that only $27$ were observed suggests that $\approx97\%$ of all Gaia NSs within $2$~kpc remain undiscovered.

We convert $N_{2~{\rm kpc}}$ to a local space density ($n$) using the effective volume in the solar neighborhood ($V_{\rm eff}$; Equation \ref{eq:v_eff_local}), where $n=N_{2~{\rm kpc}}/V_{\rm eff}(2~{\rm kpc})$. 
Our model predicts an intrinsic space density of $n=119^{+48}_{-33}~{\rm kpc^{-3}}$ for Gaia NS binaries. The true space density of Gaia NSs allows them to be interpreted in the context of other NS populations, which we detail in Section \ref{sec:ns_population_comparison}.

\subsection{Comparison to WD--MS binaries}\label{subsec:comp_to_WDMS}

\begin{figure*}
    \centering
    \includegraphics[width=0.98\linewidth]{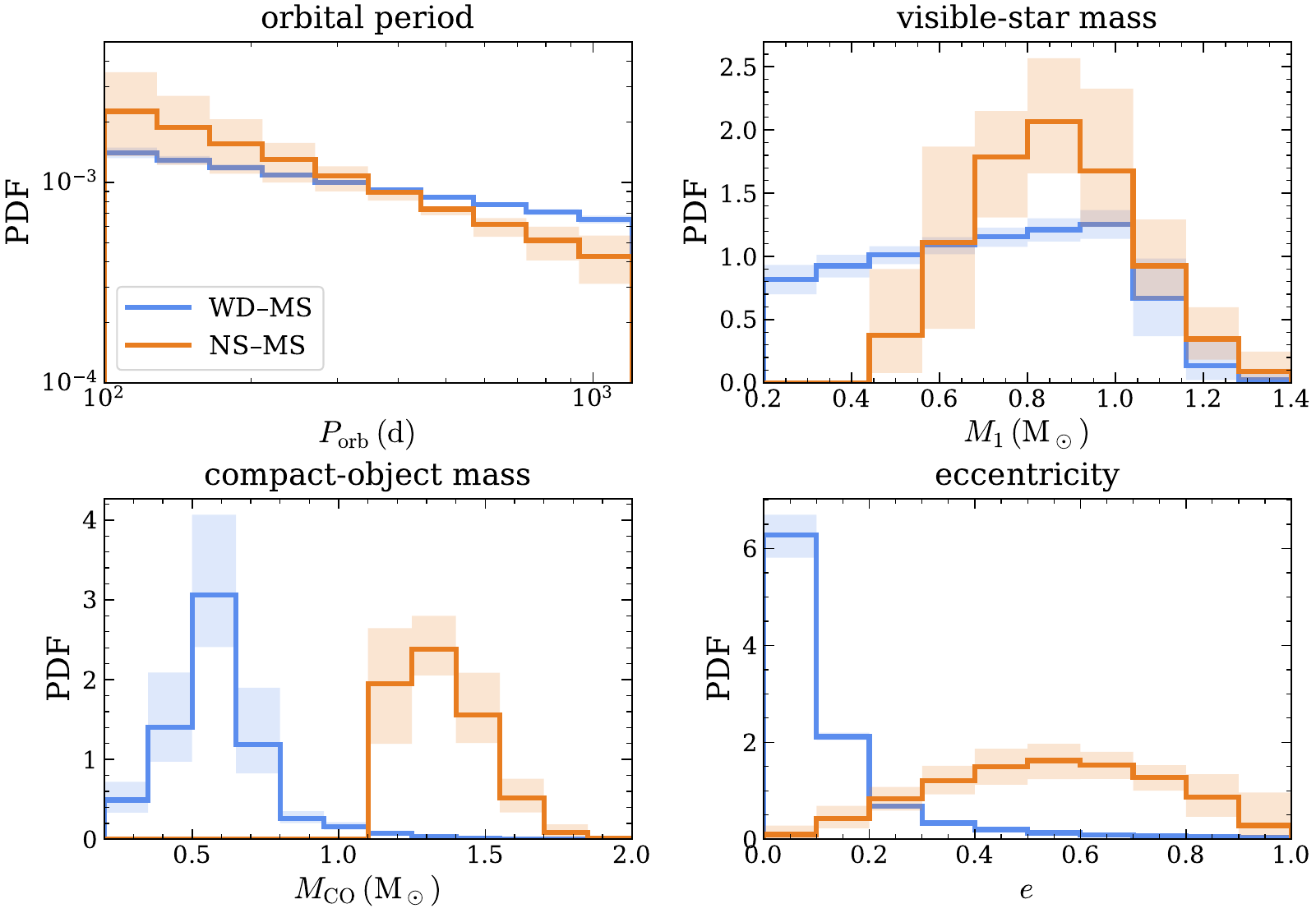}
    \caption{Comparing intrinsic distributions of au-scale WD--MS binaries (blue; \citealt{Yamaguchi2026inprep}) and NS--MS binaries  (orange; this work). Solid histograms show the posterior median probability densities, while shaded regions show the corresponding 16th--84th percentile intervals. The two populations have similar declining orbital period distributions and peaked visible-star masses, while WD--MS binaries are concentrated at lower eccentricities than NS--MS binaries.}
    \label{fig:wdms_nsms_intrinsic_comparison}
\end{figure*}

\citet{Yamaguchi2026inprep} modeled the {\it Gaia} selection function and used effectively the same hierarchical Bayesian framework to constrain demographics of the {\it Gaia} DR3 WD--MS binary population \citep[see also][]{Yamaguchi2025}. These astrometric binaries are direct analogs of the Gaia NSs: both are in au-scale orbits with FGK companions and dark compact objects. The {\it Gaia} WD--MS binaries are likely products of dynamically stable mass transfer\footnote{Common envelope evolution on the TP-AGB may also produce such wide systems \citep[e.g.,][]{Yamaguchi2024b,Belloni2024}.} from AGB donors \citep[e.g.,][]{Hallakoun2023,Garbutt2024,Shahaf2025,Yamaguchi2025,Bhattacharjee2026}.
The formation of Gaia NSs is much less constrained, but involves the additional effects of massive-star evolution, supernova mass loss, and natal kicks.
\citet{Yamaguchi2026inprep} model $P_{\rm orb}$ with a single power law, $M_1$ and $e$ with broken power laws, and $M_{\rm WD}$ with a broken power law plus a B-spline residual.

Figure~\ref{fig:wdms_nsms_intrinsic_comparison} compares the inferred intrinsic orbital-period, visible-star-mass, compact-object-mass, and eccentricity distributions of the two populations.
Over $50<P_{\rm orb}/{\rm d}<1200$, \citet{Yamaguchi2026inprep} infer a shallow decreasing period distribution, $\alpha_P=-0.445\pm0.045$, for WD--MS binaries.  Our Gaia NS fit over $100<P_{\rm orb}/{\rm d}<1600$ gives a steeper but statistically consistent slope, $\alpha_P=-0.75^{+0.37}_{-0.34}$. Interestingly, both populations prefer a decreasing period distribution with similar power-law slopes.
A larger Gaia NS sample in {\it Gaia} DR4 will provide a more constrained measurement of $\alpha_P$ while probing a larger period range, at which point a comparison with the WD--MS population will be more informative. 
If the Gaia NS period distribution, among other properties, remains similar to the WD--MS population in {\it Gaia} DR4, this would support a scenario where both populations share a similar binary mass transfer history, even though Gaia NSs also presumably survived a supernova.

The distribution of visible companions in both populations peaks near $0.8$--$0.9~{\rm M_\odot}$ (upper-right panel of Figure~\ref{fig:wdms_nsms_intrinsic_comparison}).
This comparison is also mildly selection-dependent because the AMRF cuts can affect the accessible $M_1$ range differently for WDs and NSs. At fixed dark-companion mass, larger $M_1$ lowers the AMRF signal, so higher-mass visible stars generally require more massive dark companions to lie cleanly above the single-MS locus \citep{Shahaf2019,Shahaf2024}.
Still, it suffices to say that both selected parent populations are centered on long-lived, roughly solar-type companions.

The WD mass distribution peaks near $M_{\rm WD}\approx0.57~{\rm M_\odot}$, well below the NS candidate masses (lower-left panel of Figure~\ref{fig:wdms_nsms_intrinsic_comparison}).
Only $\sim2\%$ of the inferred au-scale WD population has $M_{\rm WD}>0.8~{\rm M_\odot}$.  
The rarity of massive WDs in the WD--MS population supports the NS interpretation for the bulk {\it Gaia} NS sample, although individual ultramassive-WD (or WD--WD) contaminants remain possible.

The eccentricity distribution between the two populations is noticeably different (lower-right panel of Figure~\ref{fig:wdms_nsms_intrinsic_comparison}).  \citet{Yamaguchi2026inprep} find a WD--MS eccentricity distribution that peaks at $e=0.0768\pm0.0039$, favoring near-circular (but notably not exactly circular) orbits.  These non-zero eccentricities, and the fact that several WD--MS binaries have $e\gtrsim0.3$, suggest that AGB mass transfer can leave residual eccentricity in au-scale WD--MS binaries. The source of this eccentricity remains an active area of debate (e.g., see \citealt{vanSon2026} for a recent discussion). 
Gaia NSs, on the other hand, are much more eccentric than WD--MS binaries at the same periods and companion masses. The intrinsic eccentricity distribution of Gaia NSs is centered near $\mu_e=0.54^{+0.06}_{-0.05}$ with little support at circular orbits. The $e \gtrsim0.7$ Gaia NSs remain unconstrained due to poor {\it Gaia} sensitivity in that region (Figure \ref{fig:selection_function}), but mass loss and natal kicks would suggest that this region is also sufficiently populated.

Gaia NSs are roughly $250\times$ less common than their WD--MS analogs.  \citet{Yamaguchi2026inprep} infer a midplane density of $3\times10^4~{\rm kpc^{-3}}$ for au-scale WD--MS binaries, compared to $119^{+48}_{-33}~{\rm kpc^{-3}}$ for Gaia NSs from this work.
In other words, roughly $\sim0.4\%$ of solar-type stars host WDs in au-scale orbits \citep{Yamaguchi2025} while $\sim0.0016\%$ of solar-type stars host NSs in au-scale orbits.

WD--MS binaries show that au-scale post-mass-transfer binaries with low-mass companions are common, and that mass transfer alone can leave small but nonzero eccentricities.  Gaia NSs occupy a similar period and companion-mass regime, but are $\approx250$ times rarer and much more eccentric.  {\it Gaia} DR4 will make this comparison much sharper, where the Gaia NS population size will become $\sim10\times$ larger (Section \ref{subsec:dr4_forecasts}) and uncertainties will decrease.
Shared features may point to similar mass-transfer histories before compact-object formation, while the differences could isolate the additional effects of core collapse, natal kicks, post-supernova binary survival, and more extreme initial mass ratios.

\subsection{{\it Gaia} DR4 Forecasts}\label{subsec:dr4_forecasts}

\begin{figure*}
    \centering
    \includegraphics[width=\linewidth]{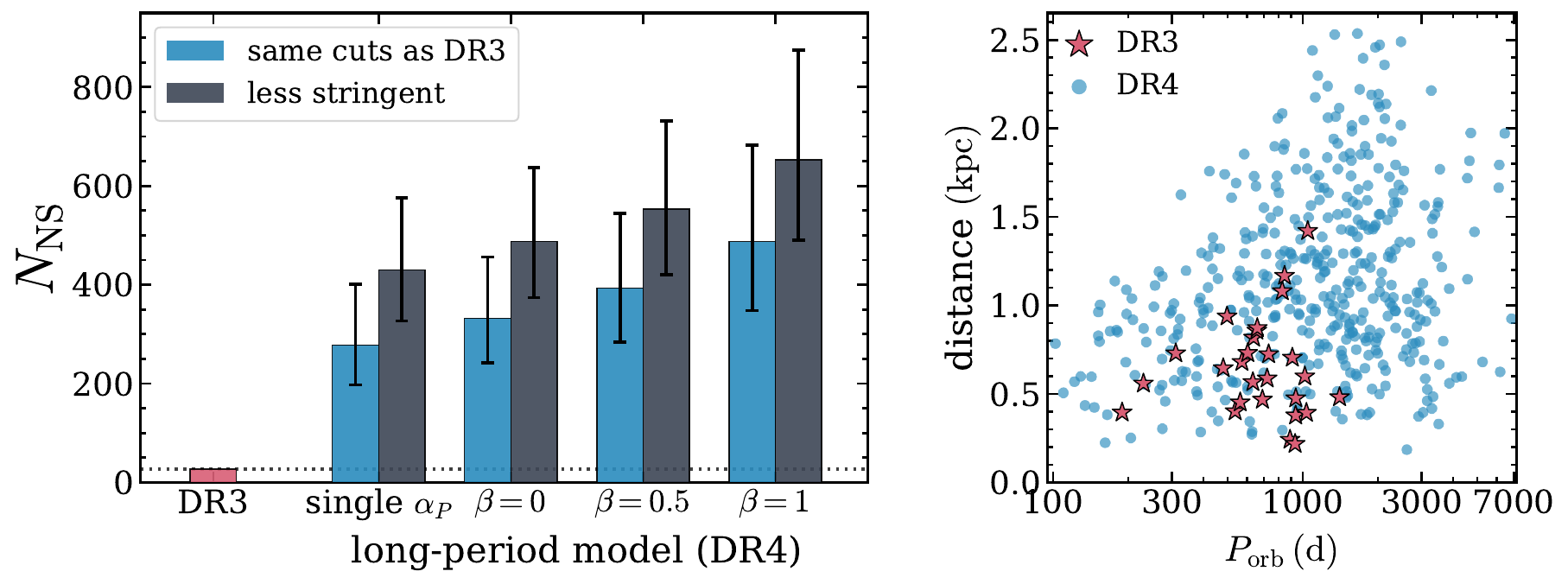}
    \caption{Forecasted yields of NS--MS binaries in {\it Gaia} DR4. 
    {\bf Left:} the number of neutron stars ($N_{\rm NS}$) 
    observed in DR3 and predicted in DR4 by our model. The DR4 predictions show results for varying assumptions about the $P_{\rm orb}$ distribution at long periods.   
    The `single-$\alpha_P$' model assumes that the same $P_{\rm orb}$ power law in DR3 extends to longer periods.  The other models keep the single $\alpha_P$ at $P_{\rm orb}<1600\,{\rm d}$ but vary the long-period tail with power-law slopes $\beta=0$, $0.5$, and $1$ (see Equation \ref{eq:dr4_power_law}). Blue bars apply the same astrometric-quality cuts used for DR3, while gray bars use less stringent criteria (see text).  
    {\bf Right:} periods and distances of the observed DR3 systems compared to predictions in DR4 from the single-$\alpha_P$ model.  {\it Gaia} DR4 will increase the detached NS--MS sample by over an order of magnitude to $\sim300$ -- $650$ systems, extending them to larger periods and distances.}
    \label{fig:dr4_broken_tail_forecast}
\end{figure*}

Using the inferred model for the underlying Gaia NS population, we predict how many Gaia NS binaries will be discovered in {\it Gaia} DR4.  
{\it Gaia} DR3 was the first release to publish non-single-star orbital solutions, but the current NS sample is shaped by the $34$-month observing baseline and astrometric-orbit quality cuts, many of which are relatively conservative \citep{GaiaCollaboration2023,Halbwachs2023}.  {\it Gaia} DR4 will be based on $66$ months of data with improved photometric and astrometric precision. Imperatively, {\it Gaia} DR4 will also include epoch astrometry, allowing for closer inspection of the astrometric orbits. All these factors will improve sensitivity to longer periods, fainter sources, and larger distances. The subset with luminous stars brighter than $G\simeq14$~mag will also have usable epoch radial velocities from Gaia RVS spectra \citep[e.g.,][]{Katz2023, Sartoretti2023}. 
The properties of Gaia NS binaries in DR4 will test major uncertainties left by DR3, including the behavior of the period distribution beyond $P_{\rm orb}\sim1000$~d and properties of the luminous companions. The long-period tail of Gaia NSs is particularly important for testing binary evolution models and prescriptions for NS natal kicks.

In this section, we forecast the yield of NS binaries in {\it Gaia} DR4. Because our DR3-calibrated model is constrained only out to $P_{\rm orb}=1600$~d, we extrapolate the period distribution to $P_{\rm orb}=7000$~d to match the longer DR4 baseline. The long-period tail is currently unconstrained, so we consider four simple models: one extends the DR3-inferred single power law over the full period range, while the other three keep the DR3-inferred slope for $P_{\rm orb}\le1600$~d and adopt independent long-period slopes of $\beta=0$, $0.5$, and $1$ for $1600<P_{\rm orb}/{\rm d}\le7000$ (Equation~\ref{eq:dr4_power_law}). These choices span declining, flat, and rising long-period distributions. In all cases, the visible-star mass, NS mass, and eccentricity distributions are drawn from the DR3-inferred posterior (Figure~\ref{fig:best_fit_distributions}). The quoted uncertainties on DR4 yields include both uncertainties from the posteriors and in the selection-function model, but the dominant systematic uncertainty is the unknown behavior of the period distribution beyond $P_{\rm orb}=1600$~d. More details on DR4 forecasting are provided in Section~\ref{subsec:dr4_forecast_inputs}.

Figure~\ref{fig:dr4_broken_tail_forecast} shows our DR4 forecasts. The left panel shows the observed DR3 count and predicted DR4 yields, while the right panel shows their orbital periods and distances. As a posterior predictive check, we estimate $26.6^{+5.5}_{-4.9}$ observed systems in DR3 when sampling from the posteriors, consistent with the $N=27$ published candidates.  
The other categories show the predicted yields in {\it Gaia} DR4.
For the single-$\alpha_P$ extrapolation, DR4 is predicted to recover $278^{+123}_{-81}$ systems assuming strict DR3-style cuts and $430^{+146}_{-103}$ systems under the less stringent cuts.  
The right panel shows a representative DR4-detected sample for this single-$\alpha_P$ extrapolation, revealing that DR4 reaches systems with $P_{\rm orb}$ of several thousand days within the local 2 kpc volume. 
Most of the newly detected systems ($60\%$) in DR4 have $P_{\rm orb}>1000$~days. The most distant system lies at $d=2.53$~kpc, reflecting the rapid loss of astrometric sensitivity with apparent magnitude. Thus, even when simply extrapolating the {\it decreasing} period distribution from DR3, DR4 is expected to increase the Gaia NS sample by $10$ -- $16\times$, with most detections residing at long periods (Figure~\ref{fig:dr4_broken_tail_forecast}).

The models with flat or increasing long-period distributions substantially boost the expected number of au-scale NS--MS binaries in DR4. With the same DR3-style cuts, the predicted DR4 yield rises from $332^{+123}_{-90}$ systems for $\beta=0$ to $487^{+196}_{-139}$ systems for $\beta=1$.  With the less stringent DR4 selection, the corresponding range is $488^{+149}_{-114}$ to $653^{+223}_{-163}$ systems.  Compared to the detached au-scale BHs expected in {\it Gaia} DR4 (`Gaia BHs'), the Gaia NS population will be substantially larger \citep{Nagarajan2025}, as was the case in {\it Gaia} DR3 ($N_{\rm BH}=2$ compared to $N_{\rm NS}=27$).
In summary, {\it Gaia} DR4 is expected to increase the detached NS--MS candidate sample by an order of magnitude, with typical yields in the range $\sim300$ -- $650$, depending mostly on the period distribution at $P_{\rm orb} \gtrsim1000$~d.

Although the long-period tail of Gaia NSs is presently unconstrained, other NS populations suggest that it is likely well-populated. Gaia NSs will eventually evolve into symbiotic X-ray binaries (SyXBs), once the present-day MS companion ascends the giant branch and the NS accretes from its wind \citep[e.g.,][]{Yungelson2019}. This X-ray-bright phase is short-lived compared to the detached NS lifetime, typically $\sim10^6$~yr, so only a handful of SyXBs are known in the Galaxy \citep{Bahramian2023,Nagarajan2024}. Curiously, the closest confirmed SyXB (4U~1700+24, $d=540$~pc) has an orbital period $P_{\rm orb}=4391$ d \citep{Hinkle2019}. Its progenitor would have been a detached NS--MS binary at a similarly wide, or somewhat shorter, orbit that is well beyond the DR3 detection limit but potentially accessible to {\it Gaia} DR4. Hence, it stands to reason that long-period Gaia NSs exist and may even be abundant. {\it Gaia} DR4 is ripe for discovering and characterizing this population.

\section{Neutron Star Masses}\label{sec:demographics_masses}

NS masses probe the physics of core-collapse, and most precise NS mass measurements are derived from recycled pulsars or X-ray binaries, where accretion can increase the observed mass relative to its birth mass \citep{Timmes1996,OzelFreire2016,Antoniadis2016,Alsing2018,You2025}. Gaia NS candidates provide a complementary, non-accreting sample. Their masses are dynamically measured from joint {\it Gaia} astrometry+RVs, and their detached orbits imply little subsequent mass growth \citep{ElBadry2024b}. 
Thus, their present-day masses are close to their birth masses.
In this section, we compare the intrinsic NS mass distribution in Gaia NSs to other NS populations.

Previous studies have found that NS mass distributions differ across populations.  Double neutron stars (DNSs) have masses concentrated near $\approx1.3~{\rm M_\odot}$, while more strongly recycled NSs extend to higher masses, likely reflecting accretion after birth \citep{Ozel2012,Kiziltan2013,Antoniadis2016,Alsing2018,Farrow2019}.  
\citet{You2025} compile dynamical NS mass measurements from several binary populations and infer the corresponding birth-mass distribution by accounting for the mass accreted by recycled pulsars.
They infer a birth-mass distribution that begins at $1.1~{\rm M_\odot}$, peaks near $\approx1.27~{\rm M_\odot}$, and declines steeply toward higher masses\footnote{In fact, redback pulsars appear systematically massive, with a median inferred NS mass of $1.78\pm0.09~{\rm M_\odot}$ \citep{Strader2019}.}.
The decline towards higher masses, they emphasize, is steep but not necessarily a hard cutoff. Namely, several dynamically measured NSs have masses $>1.5~{\rm M_\odot}$ in systems where substantial accretion is unlikely, showing that the NS birth-mass distribution has a genuine high-mass tail \citep[e.g.,][]{Barziv2001, Rawls2011,Martinez2015,Tauris2017}.

\citet{SchiebelbeinZwack2026} recently compared the original $21$ observed Gaia NS masses to the first-born NS masses in $13$ Galactic DNS systems. 
The recycled DNS component is chosen for comparison because, like the NSs in Gaia NS binaries, it usually formed first and underwent little subsequent accretion \citep[e.g.,][]{Tauris2017,Farrow2019}.
Their study shows that the observed {\it Gaia} and recycled-DNS mass distributions are similar, with a narrow component near $1.3~{\rm M_\odot}$ and a broader high-mass tail \citep{SchiebelbeinZwack2026}. 
However, as mentioned in their work, the comparison uses the {\it observed} Gaia NS sample without correcting for selection biases.
Here, we perform this comparison using the bias-corrected Gaia NS mass distribution.

We evaluate the normalized $M_{\rm NS}$ distribution over $1.1<M_{\rm NS}/{\rm M_\odot}<2.2$ by sampling posterior draws of our hierarchical model.  For the DNS sample, we use the data compiled by \citet{SchiebelbeinZwack2026} and refit their two-Gaussian model with the same uniform priors. This reproduces the DNS distribution reported in their study. We note that the recycled-DNS sample is not corrected for selection biases, but it is unclear whether these biases are strongly mass-dependent \citep{Farrow2019}.

\begin{figure}
    \centering
    \includegraphics[width=0.98\columnwidth]{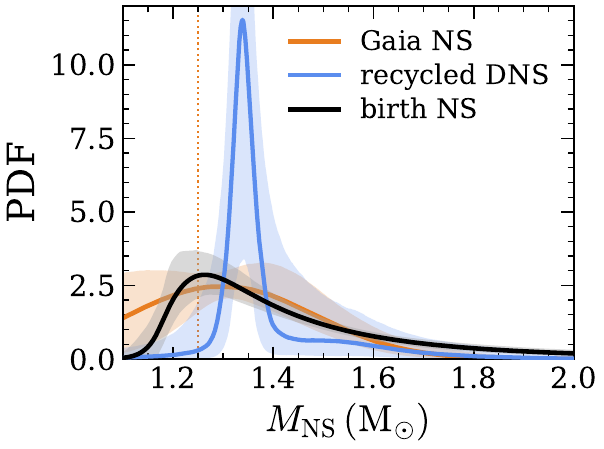}
    \caption{Neutron-star mass distributions in different populations.
    We compare the selection-corrected Gaia NS mass distribution (orange, this work) to the recycled Galactic DNS mass distribution from \citet{SchiebelbeinZwack2026} (blue) and the inferred NS birth-mass distribution from \citet{You2025} (black).  Shaded regions show the corresponding 90\% credible intervals. The dotted line marks the $1.25~{\rm M_\odot}$ selection threshold for the published Gaia NS sample, below which constraints are weak. The mass distribution of Gaia NSs broadly resembles that of the birth NS population.
    }\label{fig:ns_mass_comparison}
\end{figure}

Figure~\ref{fig:ns_mass_comparison} compares the intrinsic mass distribution of Gaia NSs to the distribution of recycled Galactic DNS and NS birth masses derived from \citet{You2025}. The Gaia NSs, recycled DNSs, and the inferred birth-mass distribution all favor most NSs near $\approx1.25$ -- $1.4~{\rm M_\odot}$, with a tail to higher masses. The high-mass tail is unlikely to be produced only by post-birth accretion. Gaia NSs have not undergone significant recycling, and recycled DNSs are expected to accrete only $\lesssim0.02~{\rm M_\odot}$ during their mildly recycled phase \citep{Tauris2017,Li2021}. Thus, it is clear that some NSs are born with $M_{\rm NS}\gtrsim1.5~{\rm M_\odot}$, consistent with earlier arguments that massive NSs need not require substantial post-birth growth \citep{Timmes1996,Kiziltan2013,Antoniadis2016,You2025}. Gaia NS1, with $M_{\rm NS}=1.88~{\rm M_\odot}$, is one clear example \citep{ElBadry2024d}.
From the intrinsic {\it Gaia} distribution, we find that $70\%$, $34\%$, and $5\%$ have $M_{\rm NS}$ larger than $1.25{\rm ~M_\odot}$, $1.4{\rm ~M_\odot}$, and $1.6{\rm ~M_\odot}$.

After correcting for selection, the agreement with recycled DNS masses is less visually striking than in the raw comparison \citep{SchiebelbeinZwack2026}. Correcting for astrometric selection effects broadens the inferred intrinsic distribution and shifts it toward lower masses. 
Recall that the published Gaia NS sample contains only candidates with $M_{\rm NS}>1.25~{\rm M_\odot}$, marked by the vertical dotted line in Figure~\ref{fig:ns_mass_comparison}. Still, the model has support below $1.25~{\rm M_\odot}$ because the prior range allows $M_{\rm NS}=1.1-1.25~{\rm M_\odot}$, and when given the choice, the data favor a distribution that decreases across the observed mass range, represented by a peak near $1.3~{\rm M_\odot}$ (see Section~\ref{subsec:intrinsic_demographics}). 

Several core-collapse models predict that lower-mass NSs receive weaker natal kicks \citep[e.g.,][]{Mandel2020,Stevenson2022,Burrows2023,Karim2026}. The low masses inferred for Gaia NSs are consistent with this scenario because they must have received relatively weak kicks to remain bound in wide binaries. For comparison, the ten eclipsing HMXBs analyzed by \citet{Falanga2015} have a median NS mass of $1.57~{\rm M_\odot}$, and HMXBs have characteristic inferred natal kicks near $100~{\rm km~s^{-1}}$ \citep{Fortin2022}. This difference may reflect a larger contribution from low-mass, low-kick explosions to the Gaia NS population. However, the HMXB sample is small, and contamination from ultramassive WDs or unresolved WD+WD binaries prevents strong conclusions about the Gaia NS mass distribution near its lower boundary.

{\it Gaia} DR4 will increase the wide NS candidate sample by an order of magnitude and extend sensitivity to longer periods \citep{ElBadry2024,Lam2025}, while radio surveys will continue to grow the recycled-DNS sample \citep[e.g.,][]{Hallinan2019, Keane2025, Shannon2025}. 
Roman microlensing surveys will provide a complementary route to isolated NS masses without requiring a binary companion \citep{Kaczmarek2026}. Isolated radio pulsars lack precise mass measurements, while binary-selected samples can be biased if NS mass, natal kick, and binary survival are correlated \citep{Burrows2024,Kaczmarek2026}. Although individual microlensing masses will generally be less precise than pulsar-timing or joint astrometric--RV masses \citep{Lam2022,Sahu2022,Mroz2025}, combining Gaia NSs, recycled DNSs, X-ray binaries, recycled pulsars, GW NSs, and isolated microlensing NSs---with separate treatment of selection effects and accretion histories---offers a promising path toward cementing the NS birth-mass distribution.
For example, if higher-mass NSs receive stronger kicks and are therefore more likely to become isolated, the selection-corrected microlensing mass distribution may be broader or shifted toward higher masses than the Gaia NS distribution.

\section{A global view of neutron star populations}\label{sec:ns_population_comparison}

By carefully modeling the {\it Gaia} selection function and applying hierarchical Bayesian inference, we inferred the intrinsic demographics of Gaia NS binaries. We now place Gaia NSs in the broader context of NS formation and binary evolution by comparing their present-day Galactic number, observable lifetime, and birth rate with those of other NS populations.

\vspace{-0.95cm}
\smallskip
\begin{deluxetable*}{lcccc}
\tablecaption{Intrinsic Galactic population size, lifetime, and birth rate of different NS populations.  Entries give the adopted range for each quantity.  Boldface marks the two quantities used as inputs to derive the third using $N_{\rm Gal}=\mathcal{R}\tau$ (see text).  Figure~\ref{fig:ns_population_lifetimes} shows a global comparison between these population properties.
\label{tab:ns_population_context}}
\tablehead{
\colhead{NS Population} &
\colhead{Number} &
\colhead{Lifetime} &
\colhead{birth rate} &
\colhead{Refs.}
\\
\colhead{} &
\colhead{$N_{\rm Gal}$} &
\colhead{$\tau$ [yr]} &
\colhead{$\mathcal{R}$ [yr$^{-1}$]} &
\colhead{}
}
\startdata
Dormant isolated NSs & $1.5\times10^7$ -- $3.5\times10^8$ & \usedinput{$5\times10^9$ -- $1.4\times10^{10}$} & \usedinput{$3\times10^{-3}$ -- $2.5\times10^{-2}$} & [1,2] \\
Ordinary radio pulsars & $30{,}000$ -- $2.5\times10^6$ & \usedinput{$10^7$ -- $10^8$} & \usedinput{$3\times10^{-3}$ -- $2.5\times10^{-2}$} & [3--5] \\
Magnetars & \usedinput{$30$ -- $300$} & \usedinput{$10{,}000$ -- $100{,}000$} & $3\times10^{-4}$ -- $3\times10^{-2}$ & [6--8] \\
Gaia NS & \usedinput{$30{,}000$ -- $100{,}000$} & \usedinput{$1\times10^9$ -- $1.4\times10^{10}$} & $2\times10^{-6}$ -- $2\times10^{-5}$ & this work \\
Symbiotic X-ray binaries & \usedinput{$8$ -- $600$} & \usedinput{$10^6$ -- $5\times10^7$} & $1.6\times10^{-7}$ -- $6\times10^{-4}$ & [9,10] \\
Be X-ray binaries & \usedinput{$300$ -- $5{,}000$} & \usedinput{$10^7$ -- $10^8$} & $3\times10^{-6}$ -- $5\times10^{-4}$ & [11--14] \\
Supergiant HMXBs & \usedinput{$50$ -- $200$} & \usedinput{$10{,}000$ -- $10^6$} & $5\times10^{-5}$ -- $2\times10^{-2}$ & [12,14--16] \\
NS LMXBs & \usedinput{$300$ -- $30{,}000$} & \usedinput{$5\times10^8$ -- $10^{10}$} & $3\times10^{-8}$ -- $6\times10^{-5}$ & [17--20] \\
Spider pulsars & \usedinput{$300$ -- $6{,}000$} & \usedinput{$10^9$ -- $10^{10}$} & $3\times10^{-8}$ -- $6\times10^{-6}$ & [21--24] \\
Recycled pulsar+WD binaries & \usedinput{$10{,}000$ -- $50{,}000$} & \usedinput{$10^9$ -- $10^{10}$} & $10^{-6}$ -- $5\times10^{-5}$ & [25--28] \\
Galactic DNSs & \usedinput{$3{,}000$ -- $30{,}000$} & \usedinput{$10^8$ -- $10^{10}$} & $3\times10^{-7}$ -- $3\times10^{-4}$ & [29--32] \\
LVK BNS mergers & -- & -- & $6.6\times10^{-7}$ -- $2.2\times10^{-5}$ & [33] \\
LVK NSBH mergers & -- & -- & $8\times10^{-7}$ -- $7\times10^{-6}$ & [33] \\
\enddata
\tablecomments{References: [1] \citet{Rozwadowska2021}; [2] \citet{Sweeney2022}; [3] \citet{Manchester2005}; [4] \citet{FaucherGiguere2006}; [5] \citet{Lorimer2008}; [6] \citet{Olausen2014}; [7] \citet{Kaspi2017}; [8] \citet{Beniamini2019}; [9] \citet{Yungelson2019}; [10] \citet{Nagarajan2024}; [11] \citet{Reig2011}; [12] \citet{Fortin2023}; [13] \citet{Liu2006}; [14] \citet{Liu2024}; [15] \citet{Mineo2012}; [16] \citet{Hurley2000}; [17] \citet{Avakyan2023}; [18] \citet{Fortin2024}; [19] \citet{Niang2024}; [20] \citet{Maccarone2022}; [21] \citet{Roberts2013}; [22] \citet{Lorimer2013}; [23] \citet{Ginzburg2021}; [24] \citet{Koljonen2025}; [25] \citet{Bhattacharya1991}; [26] \citet{Tauris2012}; [27] \citet{Gregoire2013}; [28] \citet{Lorimer2013}; [29] \citet{Chattopadhyay2020}; [30] \citet{Pol2020}; [31] \citet{Grunthal2021}; [32] \citet{ColomBernadich2023}; [33] \citet{LIGOScientificCollaboration2025}. The LVK volumetric merger rates are converted to a merger frequency assuming the space density of Milky-Way-like galaxies, $n_{\rm MW}=0.0116~{\rm Mpc^{-3}}$.}
\end{deluxetable*}

Previous studies have compared birth rates among isolated NS populations \citep{Keane2008} and between LMXBs and recycled pulsars \citep{KulkarniNarayan1988, Maccarone2022}. Here, we extend these comparisons across both isolated and binary NS populations, including the newly measured Gaia NS population.
Gaia NSs are one of the cleanest NS populations for demographic inference, owing to the well-characterized {\it Gaia} selection function for astrometric orbits \citep[e.g.,][]{ElBadry2024,Lam2025}.
The selection function for other NS classes can also be quantified. Young radio pulsars are selected by beamed spin-powered emission, magnetars by a short-lived active phase, X-ray binaries by accretion duty cycle, and LVK binary neutron stars by the LIGO sensitivity.

For each population, we compare the total number of systems in the Galaxy ($N_{\rm Gal}$), the lifetime of the relevant observable or evolutionary phase ($\tau$), and the corresponding formation rate ($\mathcal{R}$).  These quantities are related by
\begin{equation}
    N_{\rm Gal} \approx \mathcal{R}~\tau .
    \label{eq:population_context_scaling}
\end{equation}
For each class, we adopt the two quantities that are best constrained by the literature and derive the third from Equation~\ref{eq:population_context_scaling}. 
This scaling assumes an approximately steady formation rate over the relevant lifetime.
For example, isolated NSs and radio pulsars have well-characterized birth rate $\mathcal{R}$ through the CCSNe rate and active lifetime $\tau$, so $N_{\rm Gal}$ is derived.  For the binary populations, $N_{\rm Gal}$ and $\tau$ are usually best constrained, so $\mathcal{R}$ is derived. We specifically try to avoid using the estimates from population synthesis models and instead rely mostly on empirical arguments.
The adopted ranges are intentionally broad, since the selection functions and observable lifetimes differ substantially across isolated NSs, radio pulsars, magnetars, X-ray binaries, DNSs, and LVK mergers.

We outline the assumptions and reasoning behind each estimate in Appendix~\ref{subsec:population_rate_estimates}. 
We emphasize that most estimates should be interpreted as conservative, plausible ranges rather than precise demographic measurements.  
Table~\ref{tab:ns_population_context} summarizes the adopted ranges, with boldface marking the two input quantities used for each population.  In the following Section \ref{subsec:rates_interpret}, we aggregate and compare these results to highlight connections between various NS populations. We summarize our discussion in Section \ref{subsec:rates_interpret_summary}.

\begin{figure*}
    \centering
    \includegraphics[width=0.99\linewidth]{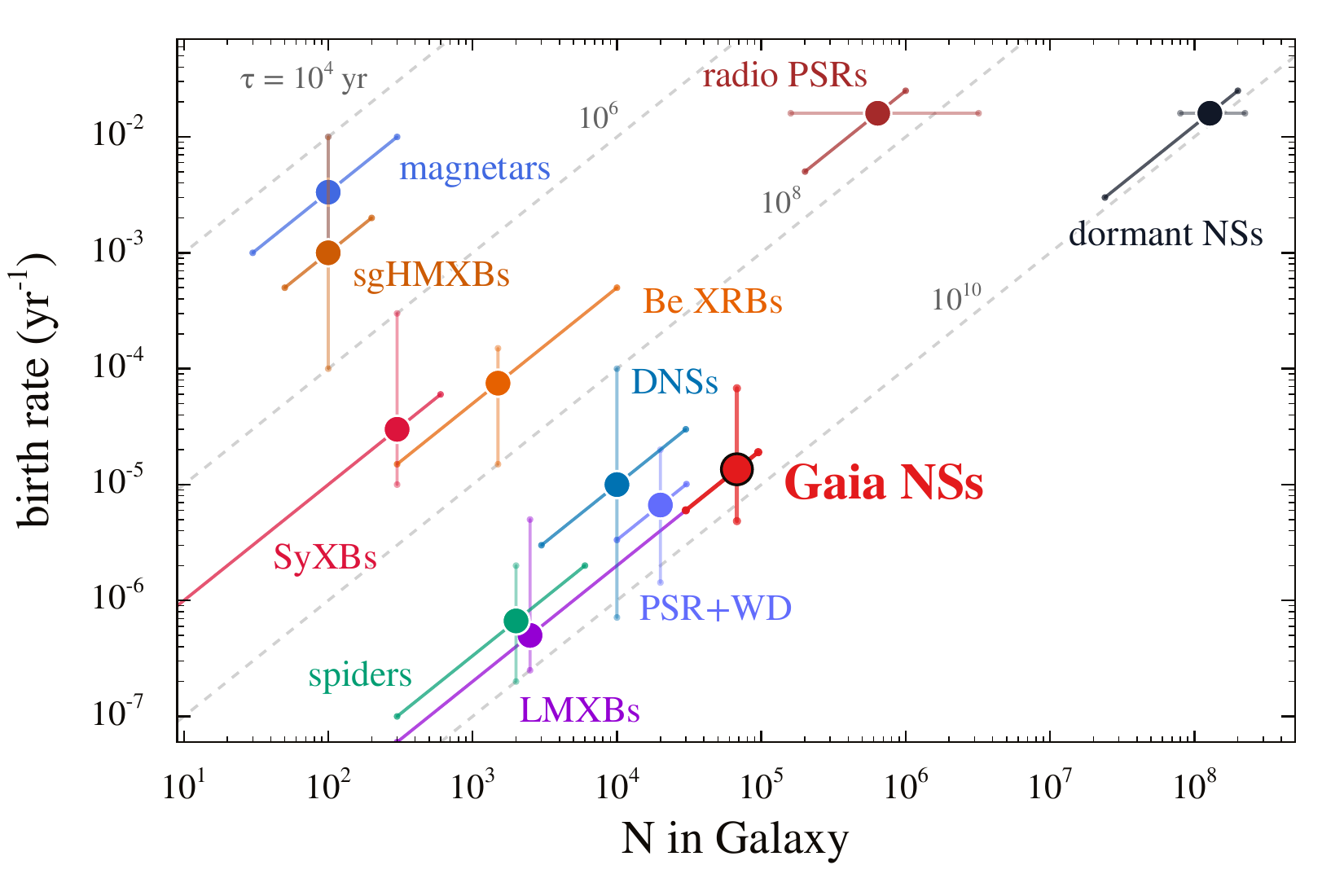}
    \caption{Intrinsic Galactic birth rate and size of NS populations.  We show the adopted values from Table~\ref{tab:ns_population_context}, with error bars showing plausible ranges. The bars are tilted because of the covariance between population size, birth rate, and lifetime. Dashed gray lines mark constant observable lifetime, $\tau=N/\mathcal{R}$, in years. The Gaia NS binaries are long-lived and relatively rare compared to all NSs formed, with $N_{\rm Gal}\sim7\times10^4$ and $\mathcal{R}\sim10^{-5}~{\rm yr^{-1}}$.}
    \label{fig:ns_population_lifetimes}
\end{figure*}

\subsection{Interpretation}\label{subsec:rates_interpret}
 
Figure~\ref{fig:ns_population_lifetimes} compares the birth rates, lifetimes, and present-day size of different NS populations. Error bars show the plausible ranges for each population, which can be tilted because population size, birth rate, and lifetime are related (Equation \ref{eq:population_context_scaling}).
The diagonal lines show constant observable lifetimes, $\tau=N_{\rm Gal}/\mathcal{R}$.
Isolated NS populations are the most common and set the relevant scale for interpreting the other rates.  Dormant isolated NSs and ordinary radio pulsars have birth rates near the Galactic core-collapse supernova (CCSNe) rate, as expected if most CCSNe produce isolated NSs and a large fraction of them pass through a radio-loud rotation-powered phase \citep{Rozwadowska2021,FaucherGiguere2006,Lorimer2008,Sweeney2022}.  Magnetars are rarer, with inferred birth rates of order $\sim10^{-3}~{\rm yr^{-1}}$, consistent with estimates that $\sim10\%$ of CCSNe produce magnetars \citep{Olausen2014,Kaspi2017,Beniamini2019}.  

\subsubsection{Massive companions and double-compact objects}\label{subsubsec:rates_interpret_highmass}

Be XRBs and sgHMXBs have the largest birth rates among the accreting populations, $\mathcal{R}\sim10^{-4}$ -- $10^{-3}~{\rm yr^{-1}}$ \citep{Reig2011,Fortin2023,Liu2024}, at least $\sim10\times$ larger than accreting NSs with lower-mass companions (e.g., LMXBs).
Massive companions are expected to be more common since such binaries are more likely to survive the first mass-transfer phase and natal kick. The initial mass-ratio distribution of OB binaries also favors massive companions at close separations \citep[$P_{\rm orb}\lesssim100~{\rm d}$;][]{Moe2013,Sana2012,Moe2015,Moe2017}.
Wide Be+NS binaries are a useful high-mass analog of Gaia NSs: both contain a post-supernova NS with a non-degenerate stellar companion at relatively wide separations, but they occupy opposite ends of the mass-ratio distribution. Comparing the two populations can isolate how NS--MS binary survival changes from the $q\sim0.05$ -- $0.1$ to $q\sim1$.

Only a small fraction of HMXBs evolve into double-compact-object systems.
Roughly $1\%$ -- $10\%$ of HMXBs are expected to become DNS binaries (Figure \ref{fig:ns_population_lifetimes}; Appendix \ref{subsubsec:rates_doubleNS}), so the Galactic DNS population constrains survival through a second mass-transfer phase and supernova.
The subset of DNSs that merge within a Hubble time has historically anchored predictions for the BNS gravitational-wave merger rate \citep{Narayan1991,Phinney1991,Kim2003,Burgay2003,Kalogera2004,Kim2006,Pol2020a,Grunthal2021}.
The current LVK BNS rate is lower than most Galactic radio-DNS extrapolations, a tension discussed recently in the context of GWTC-4, short GRBs, and $r$-process production \citep{LIGOScientificCollaboration2025,Fishbach2026}.
Also, no radio PSR--BH binaries have been confirmed compared to $\sim15$ PSR--NS, although this is not yet in strong tension with LVK NSBH merger-rate constraints \citep[e.g.,][]{FaucherGiguere2011,Chattopadhyay2020, LIGOScientificCollaboration2025}.

\subsubsection{Wide low-mass companions: Gaia NSs, SyXBs, and NS--WD descendants}\label{subsubsec:rates_interpret_lowmass}

Roughly $1$ in every $10^3$ -- $10^4$ NSs in the Galaxy reside in an au-scale orbit with a solar-type companion, corresponding to $N_{\rm Gal}\approx70{,}000$ Gaia NS-like systems (Appendix~\ref{subsubsec:rates_gaiaNS}). NS LMXBs have similar donor masses, but occupy much tighter present-day orbits and likely trace a different outcome of extreme-mass-ratio evolution. Both populations likely began with a massive NS progenitor interacting with an FGK companion. The key difference is that Gaia NSs survived as detached au-scale binaries, while NS LMXBs survived but after significant orbital shrinkage. Their relative rates, abundances, and orbital properties therefore constrain the branching between wide detached survivors and compact accreting descendants after extreme-$q$ binary interaction.
Intermediate-mass stripped-star binaries with unseen companions may provide a view of the preceding evolutionary stage \citep{Laroche2026}.

The post-mass-transfer WD+MS population provides a useful analogy. Comparing the demographics of compact WD+MS binaries to wider au-scale WD+MS binaries enables global constraints on the branching between unstable common envelope evolution and stable mass transfer \citep{Yamaguchi2025, Shariat2026}. Gaia NSs and NS LMXBs could eventually provide similar global constraints for post-mass-transfer NS binaries, both wide and compact. This comparison is currently limited by the poorly constrained NS-LMXB rate, $\mathcal{R}\sim10^{-7}$ -- $10^{-4}~{\rm yr^{-1}}$, which remains dominated by uncertain X-ray selection biases \citep{Avakyan2023,Fortin2024,Niang2024,Maccarone2022}.

SyXBs are the evolved counterparts of Gaia NS binaries. As the solar-type companion ascends the red giant branch, the NS can accrete from its wind and become an X-ray bright SyXB. The present-day Galactic SyXB population is $\sim200\times$ smaller than the Gaia NS population (Appendix \ref{subsubsec:rates_syXB}), but its X-ray-bright (observable) lifetime is also $\sim100\times$ shorter than the several-Gyr lifetime of Gaia NSs \citep[$\sim10$ Myr;][]{Nagarajan2024}, making their birth rates comparable. This is expected if Gaia NSs generally pass through a SyXB phase. The true SyXB birth rate may be higher, however, because the Gaia NS sample in Figure~\ref{fig:rate_normalization} includes only the period, eccentricity, and companion-mass range accessible to the DR3 search \citep{ElBadry2024b,ElBadry2026}.

Wider NS--MS binaries ($P_{\rm orb}\gtrsim1000$~d) can also evolve into SyXBs, but are largely absent from the {\it Gaia} DR3 sample and therefore from our Gaia NS birth-rate estimate. For example, the {\it nearest} confirmed SyXB, 4U~1700+24 ($d=0.54$~kpc), has $P_{\rm orb}=4391$~d \citep{Hinkle2019}; its detached NS--MS progenitor would have been outside the DR3 period range but should be accessible to {\it Gaia} DR4. The proximity of 4U~1700+24 suggests that such wide progenitors may be common, so the full SyXB birth rate could exceed the DR3 Gaia NS birth rate,
\begin{equation}
\mathcal{R}_{\rm SyXB} \geq \mathcal{R}_{\rm Gaia~NS}.
\end{equation}
{\it Gaia} DR4 will test this by extending astrometric sensitivity to longer orbital periods (Section~\ref{subsec:dr4_forecasts}).

After the SyXB phase, detailed binary-evolution calculations predict that all of the known Gaia NS--MS binaries will become wide PSR--WD systems \citep{Nagarajan2024,Chattopadhyay2026}.  
The amount of recycling depends on the mass-transfer physics: eccentric mass transfer concentrated near periastron produces mildly recycled pulsars with He-WD companions, while circularized Roche-lobe overflow allows longer-lived accretion, producing MSPs and, in some cases, CO-WD companions \citep{Chattopadhyay2026}.
In either case, the final orbits remain wide. Such wide orbits are underrepresented, but not absent\footnote{ Examples include PSRs J0214+5222, J0407+1607, J1711$-$4322, J1840$-$0643, J1900+0213, J1910+0423, J1917+1046, J1923+2022, J2016+1948, J2204+2700, and J0125$-$5854 \citep{Stovall2014,Lorimer2005,Lorimer2006,Knispel2013,Gonzalez2011,Martinez2019,Wang2025,Yang2026,Tan2026}.}, in the observed PSR+WD population\footnote{{\it Gaia} DR4 could also reveal a population of non-pulsar NS+WD binaries in au-scale orbits.}.

Only the subset of Gaia NSs with companions massive enough ($\gtrsim1~{\rm M_\odot}$) will have produced PSR--WD descendants by the present day ($\sim23\%$ according to our best-fit distribution; e.g., Figure \ref{fig:best_fit_distributions}).
The birth rate of this subset places a lower limit to the total formation rate of recycled pulsar--WD binaries:
\begin{equation}
    \mathcal{R}_{\rm Gaia~NS}(M_1\gtrsim1~{\rm M_\odot}) \leq \mathcal{R}_{\rm PSR+WD}.
\end{equation}
However, this is probably a conservative lower limit since the observed MSP--WD population is dominated by shorter-period, nearly circular systems \citep{Manchester2005,Lorimer2008}, whereas the Gaia NS descendants are predicted to remain at periods longer than $\sim100$~days and contribute more significantly to the mildly recycled ($P \gtrsim20$~ms) PSR population \citep{Chattopadhyay2026}.  
While the present-day birth rate of PSR+WD binaries is uncertain due to challenges of modeling their selection effects, the inequality above should remain satisfied as future efforts refine population constraints (Figure \ref{fig:ns_population_lifetimes}).

\subsubsection{NS LMXBs and spider pulsars}\label{subsubsec:rates_interpret_recycled}

NS LMXBs and spider pulsars trace the compact branch of low-mass NS binary evolution. In the standard picture, the donor star in NS LMXBs transfers mass onto the NS, spins it up, and eventually leaves behind a recycled MSP \citep[e.g.,][]{Bhattacharya1991,Tauris2012}.
Spider pulsars are one post-accretion outcome, in which the now rotation-powered NS irradiates and ablates a low-mass companion \citep{Fruchter1988,Roberts2013,Koljonen2025}. In steady state, the spider formation rate should therefore be comparable to, but not larger than, the NS-LMXB formation rate,
\begin{equation}
   \mathcal{R}_{\rm NS-LMXB} \gtrsim \mathcal{R}_{\rm spider},
\end{equation}
because only a subset of NS LMXBs are expected to become observable spider binaries. The broad empirical ranges adopted here, $\mathcal{R}\sim10^{-7}$ -- $10^{-5}~{\rm yr^{-1}}$ for both populations, satisfy this expectation within current uncertainties (Figure~\ref{fig:ns_population_lifetimes}).

Early work noted that the birthrate inferred from spider pulsars can exceed that inferred from the observed LMXB sample, implying missing or X-ray-faint accretion phases \citep{KulkarniNarayan1988,Levin2013}. More recent work argues that faint or long-recurrence-time NS LMXBs can largely resolve this tension by increasing the intrinsic LMXB population by orders of magnitude \citep{Maccarone2022,MeyerHofmeister2024}.

\subsection{Summary}\label{subsec:rates_interpret_summary}

The majority of NS progenitors (OB stars) are born in binary or higher-order systems \citep[e.g.,][]{Sana2012,Moe2017,Offner2023} and experience rich evolutionary histories. The outcomes of these complex evolutionary histories are observed today as a diverse population of NS-harboring systems throughout our Galaxy, and Figure~\ref{fig:ns_population_lifetimes} summarizes empirical constraints on their birth rates, lifetimes, and sizes. These constraints serve as checkpoints that models of NS evolution can be anchored to, within uncertainties.

Subsets of this global picture can be interpreted as continuous evolutionary sequences. 
Most NSs are born isolated, with a birth rate set by the Galactic CCSNe rate ($\sim1$ per century).
A fraction of NSs will instead remain bound to their companions, with HMXBs having the highest formation rate. As the HMXB evolves, only a small fraction of them will survive the second stage of binary interaction and the core collapse of the donor into a compact object. Those that survive this stage will become DNSs.  The subset of DNSs that merge within a Hubble time connects this Galactic evolutionary sequence to the LVK BNS merger rate.

Wide low-mass companions are rarer, but Gaia NSs now provide the first intrinsic constraints on detached, post-supernova NS binaries in au-scale orbits. They trace the wide-surviving branch of extreme-mass-ratio binary evolution, while NS LMXBs and spider pulsars trace the compact branch that likely underwent substantial orbital shrinkage during mass transfer. As Gaia NS companions evolve off the main sequence, these systems should pass through a SyXB phase and eventually become wide PSR--WD binaries, setting a lower limit on the formation rate of long-period PSR--WD systems. The relative numbers and rates of Gaia NSs, SyXBs, NS LMXBs, spiders, and PSR--WD binaries therefore constrain how often extreme-$q$ binaries survive at wide separations, shrink into compact interacting systems, or merge.

As new populations are discovered and the statistics on existing populations are refined, we anticipate significant updates to the rates and understanding of NS evolution. More uniform surveys with well-defined selection functions provide a promising avenue to constrain intrinsic demographics, shrink the uncertainties in Figure~\ref{fig:ns_population_lifetimes}, and quantify the connections between populations. 
As {\it Gaia} DR4 extends the census of au-scale and wider NS--MS binaries, uniform X-ray monitoring constrains the hidden population of faint and long-recurrence NS LMXBs, next-generation radio surveys increase the known pulsar sample by factors of several, and LVK observes more BNS/NSBH mergers, the locations and uncertainties of the populations in Figure~\ref{fig:ns_population_lifetimes} will sharpen, and a global view on NS populations will become clearer.

\section{Conclusions}\label{sec:conclusions}

The third {\it Gaia} data release revealed a new class of neutron stars (NSs) in au-scale binaries with solar-type companions \citep{ElBadry2024b,ElBadry2024d,ElBadry2026}. The discovery of these $27$ Gaia NS candidates presents a challenge for binary-evolution models, because it is unclear how these systems avoided merger or dramatic orbital shrinkage when the NS progenitors were red supergiants, and how they remained bound when the NSs formed.
A promising path forward to constrain the formation history of Gaia NSs is to characterize their intrinsic population properties: how common are Gaia NSs, and what are the distributions of orbital periods, eccentricity, main-sequence masses, and neutron star masses?

Fortunately, unlike previous samples of NSs with low-mass companions such as NS-LMXBs, the selection biases of the Gaia NS population are well-characterized and mainly governed by {\it Gaia}'s astrometric sensitivity \citep[][]{ElBadry2024}. On the other hand, contamination from ultramassive WDs or WD+WD binaries may exist and is difficult to characterize (see also Appendix~\ref{app:high_mns}).
In this study, we carefully model the {\it Gaia} selection function to determine the intrinsic properties of the Gaia NS population, including their distribution of orbital periods, eccentricities, neutron star masses, companion masses, space densities, and birth rates. Using these results, we then place the new class of Gaia NSs in the broader NS population, connecting them to isolated NSs, X-ray binaries, recycled pulsars, double neutron stars, and compact-object mergers.  Our main results are summarized as follows:

\begin{enumerate}
    \item {\it Intrinsic properties:} By modeling the {\it Gaia} DR3 selection function with {\tt gaiamock} \citep[][]{ElBadry2024} and applying a hierarchical Bayesian model, we infer the intrinsic properties of the Gaia NS population. 
    Over the range of periods, masses, and eccentricities spanned by the sample, we find (Figure \ref{fig:best_fit_distributions}):
    \begin{enumerate}
        \item a decreasing period distribution with power-law slope $\alpha_P=-0.75^{+0.37}_{-0.34}$
        \item a companion distribution peaked at $M_1\approx 0.85~{\rm M_\odot}$ ($\sigma_{M_1} \approx 0.2~{\rm M_\odot}$)
        \item a neutron star distribution peaked at $M_{\rm NS}\approx 1.3~{\rm M_\odot}$ ($\sigma_{M_{\rm NS}} \approx 0.18~{\rm M_\odot}$)
        \item an eccentricity distribution that favors moderately eccentric orbits ($e\gtrsim 0.5$).
    \end{enumerate}

    \item {\it Space density:}
    We infer an intrinsic midplane space density of $n\approx120~{\rm kpc^{-3}}$, implying that 
    $\approx1$ in $50{,}000$ solar-type stars host neutron star companions in au-scale orbits (Figure \ref{fig:rate_normalization}). 

    \item {\it Comparison to WD--MS binaries:} We compare the intrinsic demographics of Gaia NSs to the analogous au-scale WD--MS binaries from {\it Gaia} DR3 (Figure~\ref{fig:wdms_nsms_intrinsic_comparison}; Section \ref{subsec:comp_to_WDMS}).  
    Both populations have decreasing orbital-period distributions over $\sim100$ -- $1200$~days: ${\rm d}N/{\rm d}P_{\rm orb}\propto P_{\rm orb}^{-0.45\pm0.045}$ for WD--MS binaries and ${\rm d}N/{\rm d}P_{\rm orb}\propto P_{\rm orb}^{-0.75 \pm 0.35}$ for Gaia NSs. Their visible companions both peak at $M_1\approx0.8$ -- $0.9~{\rm M_\odot}$.
    The main differences are abundance and eccentricity: Gaia NSs are $\approx250\times$ less common than au-scale WD--MS binaries and favor much larger eccentricities ($e\gtrsim 0.5$ compared to $e\lesssim0.1$ for WD--MS binaries). 

    \item {\it Neutron star masses:} The inferred NS mass distribution of Gaia NSs is largely similar to the birth NS mass distribution from other NS populations \citep{You2025}. 
    It also overlaps the recycled Galactic DNS distribution, but after correcting for the Gaia selection function, the Gaia NS distribution tends to favor somewhat lower masses (Figure~\ref{fig:ns_mass_comparison}).

    \item {\it {\it Gaia} DR4 forecast:} We predict that {\it Gaia} DR4 will increase the detached NS--MS sample by over an order of magnitude, to $N_{\rm DR4}\approx300$ -- $650$ systems (Figure~\ref{fig:dr4_broken_tail_forecast}).  The largest uncertainty in the yield is the long-period tail ($P_{\rm orb}\gtrsim1600$~d), although nearby Symbiotic X-ray binaries suggest that this tail is well-populated.

    \item {\it A global view of NS populations:} We place Gaia NS binaries in the broader context of other Galactic NS populations, including isolated NSs, X-ray binaries, recycled pulsars, double NSs, and compact-object mergers (Figure~\ref{fig:ns_population_lifetimes}). Only $\approx1$ in $1000$ NSs are born into au-scale orbits with solar-type companions.  This makes the birth rate of Gaia NSs comparable to that of Galactic DNSs and about $10$ -- $100\times$ lower than that of high-mass X-ray binaries. The Gaia NS birth rate is broadly similar to the highly uncertain rates of NS LMXBs, spider binaries, millisecond pulsars, and SyXBs. Refining this empirical roadmap of NS populations with future surveys will provide a strong foundation for understanding how NS binaries form, survive, and evolve.
\end{enumerate}

\section*{Acknowledgments} \label{acknowledgments}
We thank Hans-Walter Rix, Smadar Naoz, and Sahar Shahaf for useful discussions.
C.S. acknowledges support from the Department of Energy Computational Science Graduate Fellowship.
This material is based upon work supported by the U.S. Department of Energy, Office of Science, Office of Advanced Scientific Computing Research, under Award Number DE-SC0026073. 
This research was supported in part by grant NSF PHY-2309135 to the Kavli Institute for Theoretical Physics (KITP) and in part by NSF grant AST-2540180.

The computations presented here were conducted in the Resnick High Performance Computing Center, a facility supported by Resnick Sustainability Institute at the California Institute of Technology.
This research used resources of the National Energy Research Scientific Computing Center (NERSC), a Department of Energy User Facility.

This work presents results from the European Space Agency (ESA) space mission {\it Gaia}. {\it Gaia} data are being processed by the {\it Gaia} Data Processing and Analysis Consortium (DPAC). Funding for the DPAC is provided by national institutions, in particular the institutions participating in the {\it Gaia} MultiLateral Agreement (MLA). The {\it Gaia} mission website is \url{https://www.cosmos.esa.int/Gaia}. The {\it Gaia} archive website is \url{https://archives.esac.esa.int/Gaia}.

\software{This work made use of \texttt{OverCite} \citep{Shariat2026b}, an in-editor citation tool for \LaTeX. This work made use of the following software packages: \texttt{Jupyter} \citep{2007CSE.....9c..21P,kluyver2016jupyter}, \texttt{matplotlib} \citep{Hunter:2007}, \texttt{numpy} \citep{numpy}, \texttt{pandas} \citep{mckinney-proc-scipy-2010,pandas_17806077}, \texttt{python} \citep{python}, \texttt{scipy} \citep{2020SciPy-NMeth,scipy_17467817}, \texttt{Cython} \citep{cython:2011}, \texttt{emcee} \citep{ForemanMackey2013,emcee_10996751}, \texttt{h5py} \citep{collette_python_hdf5_2014,h5py_7560547}, \texttt{schwimmbad} \citep{schwimmbad}, \texttt{seaborn} \citep{Waskom2021},  and \texttt{tqdm} \citep{tqdm_14231923}.
Software citation information aggregated using \texttt{\href{https://www.tomwagg.com/software-citation-station/}{The Software Citation Station}} \citep{software-citation-station-paper,software-citation-station-zenodo}.
}

\appendix
\twocolumngrid
\renewcommand{\theequation}{\thesection\arabic{equation}}
\renewcommand{\theHequation}{\thesection.\arabic{equation}}

\section{Corner plot}\label{app:corner_plot}

\begin{figure*}
    \centering
    \includegraphics[width=0.9\linewidth]{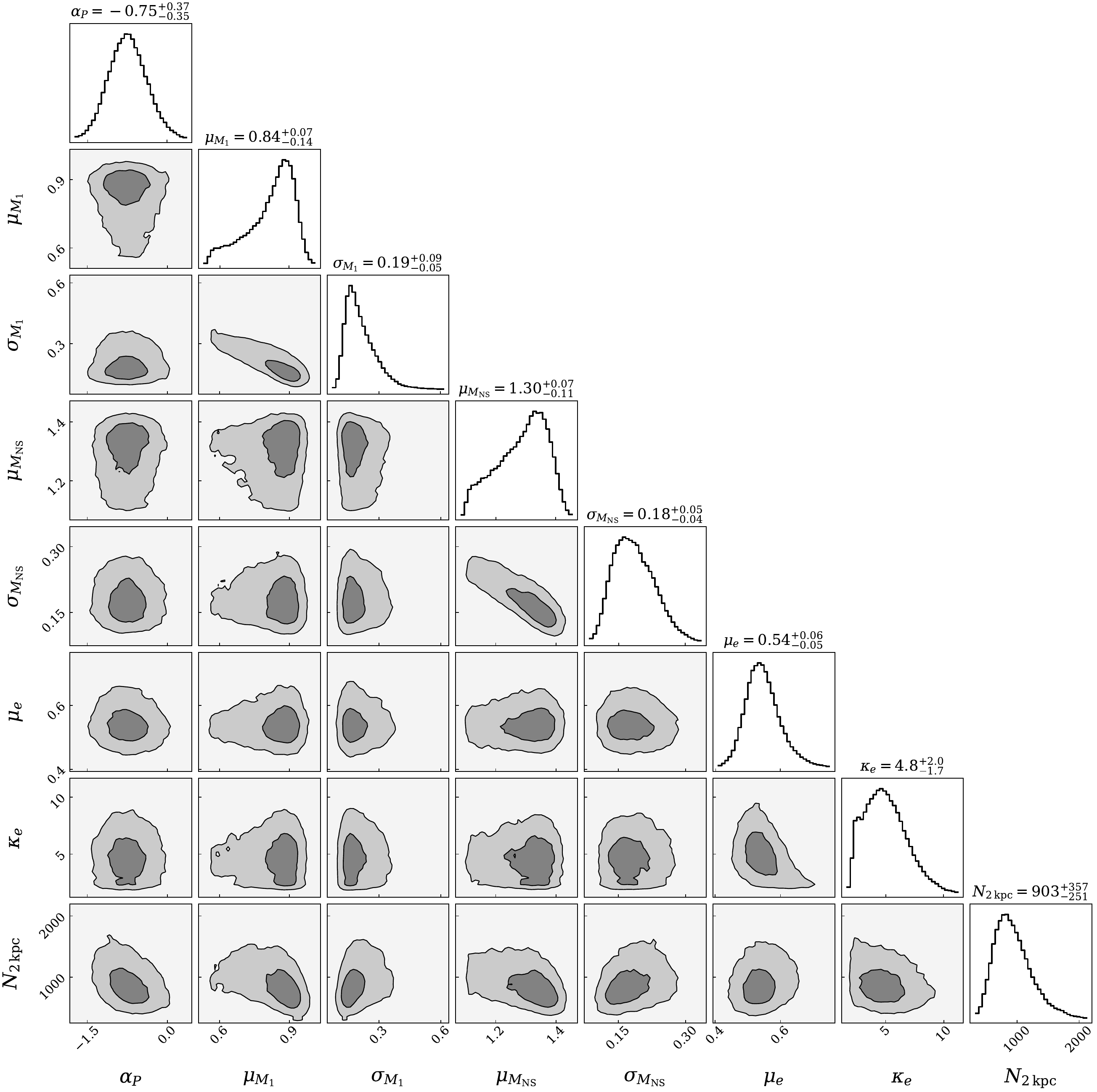}
    \caption{Posterior corner plot for the fiducial Gaia NS fit.  The plotted parameters are the period slope $\alpha_P$, visible-star mass parameters $(\mu_{M_1},\sigma_{M_1})$, NS candidate mass parameters $(\mu_{M_{\rm NS}},\sigma_{M_{\rm NS}})$, eccentricity parameters $(\mu_e,\kappa_e)$, and intrinsic count $K_{\rm 2kpc} \equiv N_{\rm 2~kpc}$.  The main demographic result is visible directly: the period slope is negative, the visible-star masses peak below a solar mass, the NS candidate masses peak near the low end of the fitted range, and the eccentricities are centered away from zero.  Filled contours enclose 1$\sigma$ and 2$\sigma$ regions.}
    \label{fig:corner}
\end{figure*}

Figure~\ref{fig:corner} shows the posterior for the fiducial fit used in the paper.  The broadest posterior direction is the usual normalization--selection degeneracy: models with a smaller effective selection fraction require a larger intrinsic population to reproduce the 27 selected systems.

\section{Population Diagnostics and Alternative Parameterizations}\label{app:population_diagnostics}

Our main analysis uses a simple separable population model in 
$x=(P_{\rm orb},M_1,M_{\rm NS},e)$.  This choice is useful for the current DR3 sample because only $27$ systems are available, but it also means that the inferred 1D distributions can depend on the assumed functional forms of the underlying distributions.  Here, we 
repeat the hierarchical Bayesian inference with more flexible parameterizations. 
We identify that the main conclusions presented in this work remain mostly insensitive to the particular choice of parameterization. Namely, the intrinsic Gaia NS population consistently prefers a decreasing power-law slope, and the inferred size of the population is stable to choices in the demographic model.

The first alternative is a broken-power-law model, which replaces the truncated-normal mass distributions and the Beta eccentricity distribution with one-break power laws.
The second alternative starts from the same broken-power-law model and adds a smooth spline residual to the $M_{\rm NS}$ distribution. These alternatives are motivated by \citet{Yamaguchi2026inprep}, who use them to constrain the WDMS population. Both alternatives use the same observed sample, injection archive, selection flag, and likelihood as the fiducial model.  The period distribution is still modeled as a power law, so $\alpha_P$ can be compared directly across the runs.

\begin{figure}
    \centering
    \includegraphics[width=0.85\columnwidth]{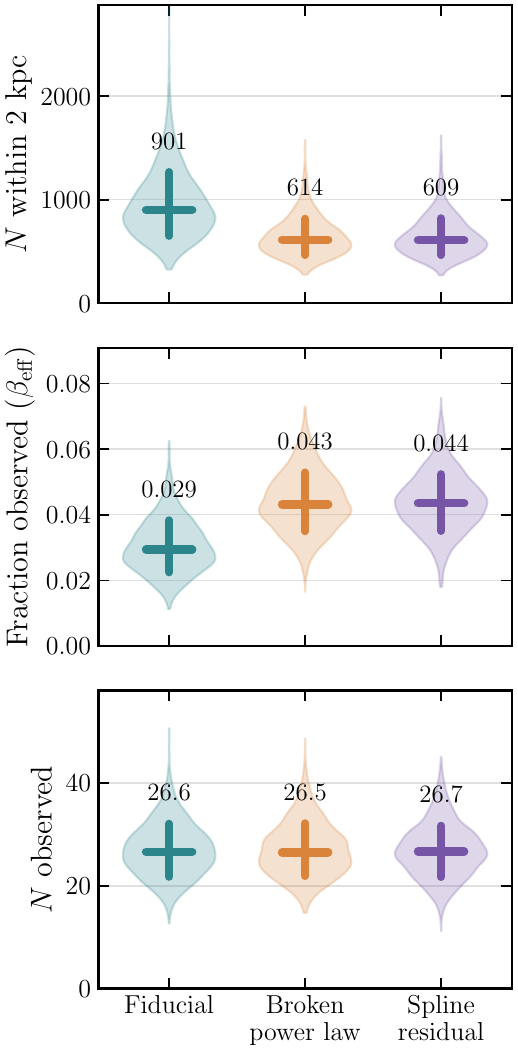}
    \caption{Effect of changing the parameterization.  Each model is fit to the same 27 Gaia NS candidates with the same selection function.  The bottom panel shows that all accepted models reproduce the observed count well, while the middle panel shows the corresponding selected fraction.  A larger selection fraction implies a smaller intrinsic count (top panel).  
    All three models reproduce the observed count, while the flexible models favor a larger effective selection fraction and a smaller inferred number within 2 kpc.
    The estimates are each consistent within $1\sigma$.}
    \label{fig:alt_model_rates_app}
\end{figure}

The alternative models give similar selected-sample predictions. Figure \ref{fig:alt_model_rates_app} shows the predicted rates and counts of Gaia NSs under the varying model assumptions.  The fiducial model estimates $K_{\rm 2kpc}=901^{+366}_{-251}$
and $\beta_{\rm eff}=0.0294^{+0.0089}_{-0.0070}$, giving an
expected count in DR3 of $26.6^{+5.5}_{-4.9}$.  
The broken-power-law model gives $N_{\rm 2~kpc}=614^{+203}_{-149}$ and $\beta_{\rm eff}=0.0432^{+0.0097}_{-0.0082}$ and the spline-residual model gives $N_{\rm 2~kpc}=609^{+212}_{-143}$ and $\beta_{\rm eff}=0.0435^{+0.0089}_{-0.0084}$. Both predict  $\approx27$ selected systems in DR3, and these shifts are smaller than the current statistical uncertainty.

Figures \ref{fig:highm2_truncnorm} and \ref{fig:highm2_wdparam} show the inferred intrinsic distributions under the different parameterizations (black curves). As revealed, they show the same qualitative features as the fiducial model (Figure \ref{fig:best_fit_distributions}).
The period slope also remains similar across the different runs, with median $\alpha_P\approx-0.4$ to $-0.75$.  We therefore treat the declining period distribution as a robust result.

We further experimented with a model parameterized in mass ratio, $q=M_{\rm NS}/M_1$.  This model can reproduce the selected count, but the inferred marginal distributions in $M_1$ and $M_{\rm NS}$ become sensitive to the selection correction.  The reason is that with only 27 selected systems, a flexible $q$ distribution can trade off the visible-star mass function against the dark-companion mass distribution while leaving the selected sample nearly unchanged.  Similar behavior arises in the WDMS population, for which the sample is much larger \citet{Yamaguchi2026inprep}.

\section{High $M_{\rm NS}$ Subsample}\label{app:high_mns}

\begin{figure*}
    \centering
    \includegraphics[width=0.95\linewidth]{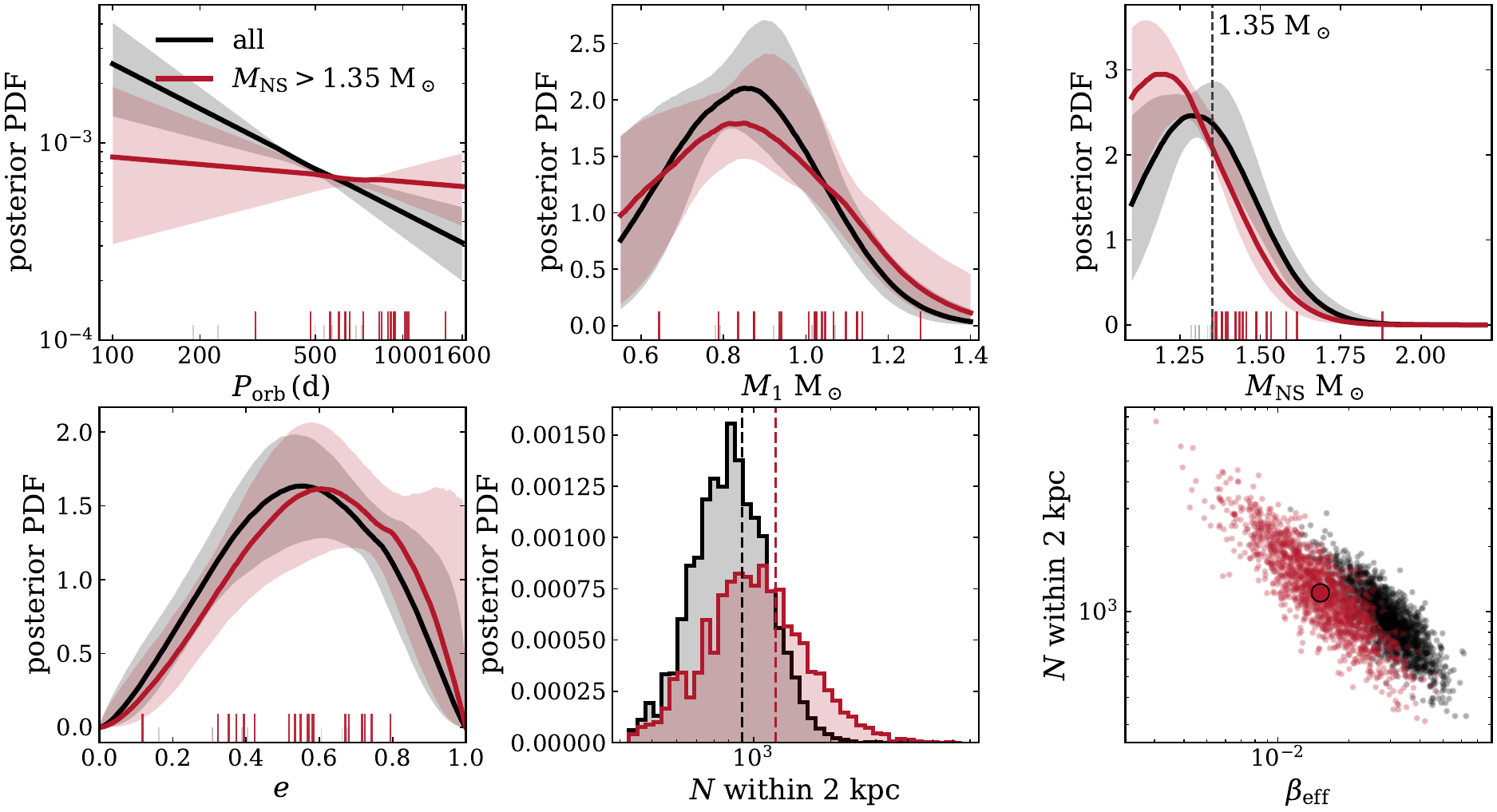}
    \caption{Population inference using the $M_{\rm NS}>1.35~{\rm M_\odot}$ subset with the fiducial parameterization. We compare the full fit (black) to the 19-object high-mass subset (red).
    The high-mass subset gives a slightly shallower period distribution, while keeping the visible-star mass and eccentricity distributions roughly the same.}
    \label{fig:highm2_truncnorm}
\end{figure*}

The main contaminant in the Gaia NS population is ultramassive WDs \citep[][]{ElBadry2024b,ElBadry2024d,ElBadry2026}.
Our main analysis treats all $27$ systems as one Gaia NS candidate population, which is the most direct interpretation of the published sample \citep[see][for a detailed discussion]{ElBadry2024b}. The lower-mass end (i.e., below the Chandrasekhar mass limit) is where ultramassive-WD contamination is possible.  
As a robustness check, we repeat the population inference using only the candidates with $M_{\rm NS}>1.35~{\rm M_\odot}$, where virtually no WDs are expected.  This cut leaves $19$ systems among the initial $27$ and removes most of the objects below the Chandrasekhar mass. Practically, this filtering changes the modeling in two ways. First, it changes the selection function to enforce $M_{\rm NS}>1.35~{\rm M_\odot}$. Second, it changes $N_{\rm det}$ to $19$ in the likelihood (Equation \ref{eq:hbi_log_likelihood}). The {\it Gaia} sensitivity remains unchanged.

\begin{figure*}
    \centering
    \includegraphics[width=0.95\linewidth]{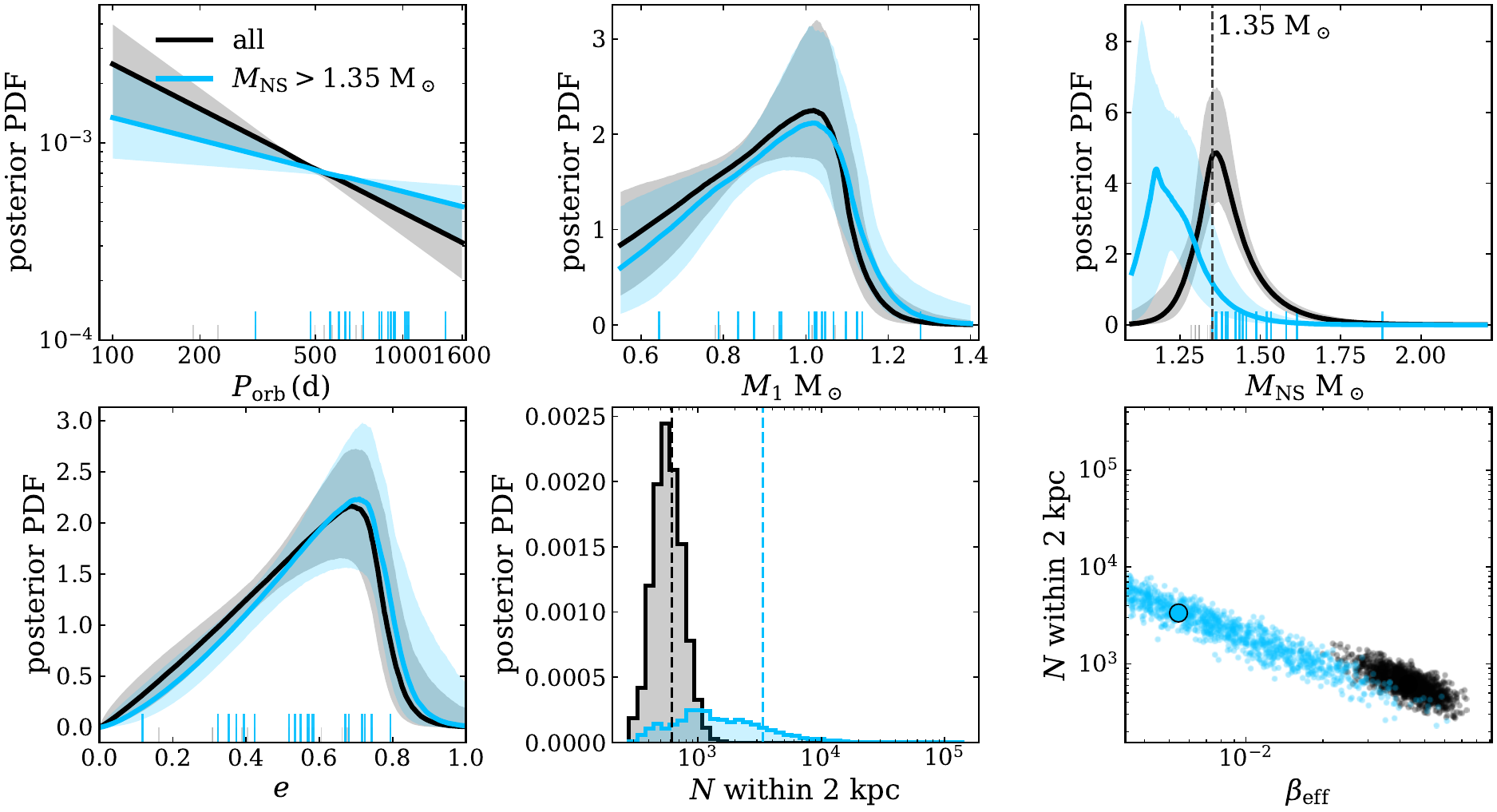}
    \caption{Same as Figure~\ref{fig:highm2_truncnorm}, but using a broken-power-law parameterization for both the full (black) and high-$M_{\rm NS}$ subset (blue).  The same pattern appears in this more flexible model. The high-mass subset gives a shallower period distribution and a broader normalization posterior, while the visible-star and eccentricity distributions remain consistent with the fiducial fit.}
    \label{fig:highm2_wdparam}
\end{figure*}

Figures \ref{fig:highm2_truncnorm} and \ref{fig:highm2_wdparam} show the results for the high-mass-restricted reruns (colored curves), including variations to the parameterization (see Appendix \ref{app:population_diagnostics}). Figure \ref{fig:highm2_truncnorm} assumes the fiducial truncated normal distributions for the masses, and Figure \ref{fig:highm2_wdparam} assumes a broken power law for the masses and eccentricity.

In the truncated-normal model, the high-$M_{\rm NS}$ subset gives $K_{\rm 2kpc}=1221^{+742}_{-430}$, $\beta_{\rm eff}=0.0151^{+0.0071}_{-0.0050}$, and an expected selected count $\mu=18.8^{+4.4}_{-4.1}$.  The inferred intrinsic count is therefore consistent with the all-27 result, but with a lower effective selection fraction because the selected sample is restricted to the high-mass candidates.  The main change is in the period distribution: the subset gives $\alpha_P=-0.13^{+0.51}_{-0.46}$, compared to $\alpha_P=-0.75^{+0.37}_{-0.34}$ for the full sample.  Thus, the high-$M_{\rm NS}$ candidates still favor a flat or declining period distribution, and the strongest preference for short periods comes from considering the full sample.

We repeat the same test with a broken power law parameterization from Appendix~\ref{app:population_diagnostics}. We find that the result is roughly the same.  The high-$M_{\rm NS}$ subset gives a shallower period slope, $\alpha_P=-0.37^{+0.26}_{-0.36}$, compared to $\alpha_P=-0.75^{+0.37}_{-0.32}$ for the all-27 fit.  The normalization is less tightly constrained, with $K_{\rm 2kpc}=3349^{+7112}_{-2162}$ and $\beta_{\rm eff}=0.0054^{+0.0095}_{-0.0037}$, because the smaller selected sample and flexible mass model allow more posterior weight at low effective selection fractions.

The high-$M_{\rm NS}$ analysis therefore supports the main conclusions.  The size of the Gaia NS population remains large ($600-900$), the eccentricity distribution remains weighted toward moderate-to-large eccentricities, and the visible companions remain near solar mass.  The main quantity that changes is the period slope when fitting high-$M_{\rm NS}$ separately.

\section{Estimating formation rates}\label{subsec:population_rate_estimates}
In this section, we compile estimates for the present-day Galactic number, characteristic lifetime, and birth rate of each NS population used in Figure~\ref{fig:ns_population_lifetimes}. These estimates are intended to be conservative ranges. Table~\ref{tab:ns_population_context} summarizes the adopted values and marks which quantities are used as inputs for each population.

\subsection{Gaia NS binaries}\label{subsubsec:rates_gaiaNS}
For the Gaia NS systems, we measured the local space density (Section~\ref{subsec:intrinsic_demographics}) to be $n=119_{-33}^{+48}~{\rm kpc^{-3}}$.  Multiplying by the thin-disk effective volume (Equation \ref{eq:v_eff_galactic}) gives
$N_{\rm Gal}\approx6.8^{+2.8}_{-1.9}\times10^4$, with an additional factor-of-few systematic uncertainty from the Galactic density model and the selection-function correction.
We therefore adopt $N_{\rm Gal}\sim7\times10^4$, with a conservative range $2\times10^4$ -- $2\times10^5$.  The relevant observable lifetime is the detached lifetime of the visible star.  The hosts have $M_1\approx0.6$ -- $1.3~{\rm M_\odot}$, corresponding to main-sequence lifetimes of roughly $1$ -- $14~{\rm Gyr}$. We use this $N_{\rm Gal}$ and the detached lifetime ($\tau$) as the two inputs to derive their birth rate $\mathcal{R}$. Dividing $N_{\rm Gal}$ by $\tau$ gives $\mathcal{R}\sim2\times10^{-6}$--$10^{-4}~{\rm yr^{-1}}$.  Compared to the core-collapse supernova rate, $1$ -- $2$ per century \citep{Rozwadowska2021}, the birth rate implies that only $\sim10^{-4}$ -- $10^{-2}$ of NSs born reside in long-lived, au-scale detached NS--MS binaries with solar-type companions.

\subsection{Isolated NSs, radio pulsars, and magnetars}\label{subsubsec:rates_isolatedNS}
The vast majority of core-collapse supernovae (CCSNe) result in neutron stars, making the CCSNe rate a functional approximation for the NS birth rate.  \citet{Rozwadowska2021} infer $\mathcal{R}=1.63\pm0.46$ CCSNe per century in the Milky Way, and most of the neutron stars produced in these supernovae will be retained in the Galaxy\footnote{For the fiducial log-normal kick distribution of isolated young pulsars \citet{Disberg2025}, 
$\ln(v_{\rm kick}/{\rm km~s^{-1}})=5.60\pm0.68$, $\sim85\%$ of kicks are below the escape speed of the solar-neighborhood $v_{\rm esc}\approx550~{\rm km~s^{-1}}$. The escape fraction decreases even more towards the Galactic center.}. We use the retained-NS birth rate ($\mathcal{R}$), within $2\sigma$ of the CCSNe rate, and lifetime ($\tau$) from $5$ -- $14$ Gyr to derive the present-day number ($N_{\rm Gal}$).
Integrated over the age of the Milky Way, and allowing for natal-kick losses and the fraction of core collapses that form black holes, this implies a present-day reservoir of order $10^8$ dormant NSs \citep{Sweeney2022}.  

Ordinary (non-recycled) radio pulsars are the active, rotation-powered subset of the NS population.  The observed radio catalog contains a few thousand pulsars \citep{Manchester2005}, but the sample is heavily biased. Radio surveys miss pulsars because of beaming, luminosity limits, sky coverage, dispersion, and scattering. Since most NSs are expected to go through a radio pulsar phase, the radio pulsar birth rate is similar to the CCSNe rate, with characteristic radio-active lifetimes of $10^7$ -- $10^8~{\rm yr}$ \citep{FaucherGiguere2006}.  We use this birthrate and the radio-active lifetime to derive $N_{\rm Gal}=10^5$ -- $10^6$ active radio pulsars in our Galaxy today.

Magnetars occupy a much shorter-lived active high-energy phase.  The McGill Magnetar Catalog lists $\sim30$ magnetars and candidates \citep{Olausen2014}, setting the lower limit for our estimate.  This observed number is likely to include most of the active bursting/X-ray-bright sample, but it can miss faint, weakly bursting, or dormant magnetars \citep{Rea2011,Kaspi2017}.  
For the plotted active-magnetar population, we use the active number ($N_{\rm Gal}$) and active lifetime ($\tau$) as inputs to derive a birth rate $\mathcal{R}$. Adopting $N_{\rm Gal}\sim100$ with a conservative range $30$ -- $300$, and $\tau\sim10^4$ -- $10^5~{\rm yr}$ provides a formation rate estimate of $10^{-3}$ -- $10^{-2}~{\rm yr^{-1}}$ \citep{Kaspi2017,Beniamini2019}.  This value is consistent with previous estimates predicting $\sim10\%$ of CCSNe produce magnetars \citep[e.g.,][]{Gill2007,Keane2008, Beniamini2019}. Specifically, \citet{Beniamini2019} infer a Galactic magnetar formation rate of $2.3$ -- $20~{\rm kyr^{-1}}$ from the observed magnetar population, spin-down properties, magnetic activity, and supernova-remnant associations, consistent with our adopted $\mathcal{R}$ range.

\subsection{High-mass X-ray binaries}\label{subsubsec:rates_HMXB}

We consider two classes of actively accreting NS binaries with high-mass companions: Be X-ray binaries (BeXRBs) and supergiant high-mass X-ray binaries (SgHMXBs).
BeXRBs comprise about half of the known Galactic HMXB catalog. The Galactic HMXB catalog contains $74$ Be XRBs among $152$ known HMXBs \citep{Fortin2023,Liu2024}, providing the strict lower limit for $N_{\rm Gal}$. Because BeXRBs are transient, concentrated near the Galactic plane, and require both X-ray detection and optical/infrared counterpart classification \citep{Reig2011}, the total number in the Galaxy is expected to be larger.
The Magellanic Clouds provide a useful anchor on the total size of the BeXRB population because their distances are known, Galactic extinction is less severe, and they can be monitored in a single pointing, making their HMXB samples more complete.

The Large Magellanic Cloud (LMC) provides the best probe.  Its SFR has been relatively steady for the past several Gyr with SFR=$0.2~{\rm M_\odot}~{\rm yr^{-1}}$ \citep{HarrisZaritsky2009}, about $10\times$ smaller than the Milky Way \citep[$1.5$ -- $2~{\rm M_\odot}~{\rm yr^{-1}}$;][]{LicquiaNewman2015}, with a slightly lower metallicity \citep[$0.5\times$ solar;][]{Kaltenbrunner2026}.  Among the $40$ classified LMC HMXBs from \citet{Antoniou2016}, $33$ ($83\%$) are BeXRBs and $4$ ($10\%$) are supergiant systems, corresponding to $83\%$ and $10\%$ of the total HMXBs.  Scaling the classified LMC BeXRB sample to the Milky Way using the ratio of SFRs estimates $\sim300$ active BeXRBs in the Milky Way, before correcting for LMC incompleteness.  
XMM-Newton observations of the LMC also discovered new HMXBs, which are mainly Be systems \citep{vanJaarsveld2018,Haberl2022}. The all-LMC eROSITA catalog now lists $53$ HMXB objects/candidates, including $28$ confirmed HMXBs and $27$ new eROSITA detections with completeness for sources brighter than $L_X\sim{\rm few}\times10^{34}~{\rm erg~s^{-1}}$ \citep{Kaltenbrunner2026}.  Quiescent BeXRBs, low-duty-cycle transients, and optically ambiguous systems are still missing and not considered in scaling the LMC values to the Milky Way.

The SMC provides another reference point, but is likely more difficult to scale directly to the Milky Way. Namely, its metallicity is lower, and the HMXB population is much larger per unit mass than the Milky Way: \citet{Haberl2016} compile $148$ confirmed and candidate SMC HMXBs and identify $121$ relatively high-confidence systems, the vast majority with Be companions.  Since the Milky Way SFR is roughly $30$ -- $50$ times larger than the SMC's, a direct SMC-to-Milky Way scaling would imply several thousand BeXRBs or more.  However, it has been noted that the SMC is unusually efficient at producing BeXRBs, probably because of its recent star-formation history $\sim50$~Myr ago and low metallicity \citep{Antoniou2010,Antoniou2016,Vinciguerra2020}.  

For our estimates, we adopt a broad intrinsic Galactic number of BeXRBs $N_{\rm Gal}=3\times10^2$ -- $5\times10^3$.  The lower end is anchored by the Galactic catalog plus the conservative LMC scaling, while the upper end allows for the much larger correction due to observational incompleteness and scaling from the SMC.  The BeXRB lifetime will be roughly the main-sequence lifetime of the B-type star ($10^7$ -- $10^8~{\rm yr}$) corresponding to $\mathcal{R}\sim10^{-5}$ -- $10^{-3}~{\rm yr^{-1}}$ with a fiducial value of a few $\times10^{-4}~{\rm yr^{-1}}$.
For BeXRBs, we therefore use the inferred Galactic number and BeXRB lifetime to derive the birth rate from $\mathcal{R}=N_{\rm Gal}/\tau$.

SgHMXBs make up most of the remaining HMXB population. Among the $152$ known Galactic HMXBs \citep{Fortin2023}, $52$ are classified as SgHMXBs, including $27$ supergiant fast X-ray transient candidates and $5$ sgB[e] systems. The lower limit on the Galactic number is thus $52$.  The donor stars are luminous, making the optical counterpart classification more complete.
However, counterpart identification and X-ray discovery remain incomplete because many systems lie in crowded, highly extincted regions and transient systems can have low duty cycles \citep{Ducci2014,Fortin2023}.  We therefore adopt $N_{\rm Gal}\sim10^2$ and allow a broad range of $50$--$200$, corresponding to the observed catalog scale with up to a factor of $\approx4$ incompleteness correction.
The relevant lifetime is the time spent in the X-ray-bright sgHMXB stage. 
Population-level HMXB arguments use characteristic bright lifetimes (i.e., wind mass transfer) of order $10^5~{\rm yr}$ \citep{Mineo2012,Quast2019}, while the donor's post-main-sequence evolution gives an upper limit of order $10^6~{\rm yr}$ for the relevant OB-star masses \citep{Hurley2000}.  We therefore adopt a deliberately broad range $\tau_{\rm sgHMXB}\sim10^4-10^6~{\rm yr}$.  The implied formation rate is $5\times10^{-5}$ -- $2\times10^{-2}~{\rm yr^{-1}}$.

\subsection{Symbiotic X-ray binaries}\label{subsubsec:rates_syXB}

Symbiotic X-ray binaries (SyXBs) are binaries where a NS accretes from the wind of an evolved red-giant companion. While they comprise only a small observed sample, they provide a spotlight on the fate of the Gaia NSs. Only a handful of SyXBs are known \citep[e.g.,][]{Yungelson2019,Bahramian2023}, and \citet{Nagarajan2024} list six systems with secure or candidate SyXB classifications.  This observed number is a lower limit because SyXBs are often wind-fed and  X-ray faint compared to Roche-lobe-overflow LMXBs, and often lack measured orbital solutions due to their long periods \citep{Hinkle2006,Hinkle2019,Hinkle2024}.

We estimate the present-day Galactic SyXB population in two ways.  First, using the nearby SyXBs compiled by \citet{Nagarajan2024}, we find that there are three known systems within $3~{\rm kpc}$, with the closest one at $d\sim0.5~{\rm kpc}$.  These two estimates imply local space densities of $n=0.027~{\rm kpc^{-3}}$ ($d \leq 3~{\rm kpc}$) and $n=2.7~{\rm kpc^{-3}}$ ($d \leq 0.5~{\rm kpc}$), corresponding to $N_{\rm Gal}\sim8$ -- $600$ after multiplying by the thin-disk effective volume.  Second, \citet{Nagarajan2024} argue that IGR~J16194--2810, at $d\approx2.1~{\rm kpc}$ and $P_{\rm orb}=192.7$ d, is probably the nearest SyXB with $P_{\rm orb}\lesssim1000$~d, i.e., the subset whose orbits should be detectable by {\it Gaia} DR3.  Comparing this distance to the distance of the closest Gaia NS, $d\sim250$ pc, implies a local SyXB density roughly $\sim100$ times lower than the detached Gaia NS population, which also roughly reflects the ratio of their lifetimes.  Since we infer $n\approx120~{\rm kpc^{-3}}$ for the Gaia NS--MS binaries (Figure~\ref{fig:rate_normalization}), this scaling gives $n_{\rm SyXB}\sim1~{\rm kpc^{-3}}$, or $N_{\rm Gal}\sim3\times10^2$.  This lies within the range from the direct nearby-SyXB estimate, so we adopt $N_{\rm Gal}\sim3\times10^2$ with a broad range $8$ -- $600$.

The observable lifetime for SyXBs is relatively short compared to Gaia NSs.  \citet{Nagarajan2024} use MESA models for IGR~J16194--2810 and find that the red giant will begin Roche-lobe overflow in $5$ -- $10$ Myr. Their fiducial model enters RLO in $5.9$ Myr and remains in the high mass-transfer phase for $\approx3.6$ Myr before leaving behind a $\approx0.4~{\rm M_\odot}$ He WD in a $\sim900$ d orbit.  We therefore adopt an X-ray-bright SyXB lifetime of order $\tau\sim1$ -- $50~{\rm Myr}$ to be conservative and allow for a broad range in orbital period, donor evolutionary state, and X-ray duty cycle.  Combining $N_{\rm Gal}\sim8$ -- $600$ with this lifetime range gives $\mathcal{R}_{\rm SyXB}\approx2\times10^{-7}$ -- $6\times10^{-4}~{\rm yr^{-1}}$, with a fiducial value of $3\times10^{-5}~{\rm yr^{-1}}$.  
For SyXBs, we thus use $N_{\rm Gal}$ and the X-ray-bright SyXB lifetime $\tau$ to derive $\mathcal{R}$.

\subsection{Low-mass X-ray binaries}\label{subsubsec:rates_LMXB}

Neutron star low-mass X-ray binaries (NS LMXBs) comprise the majority of LMXBs, with a few hundred systems observed in the Galaxy \citep[e.g.,][]{Bahramian2023}.  The recent \citet{Fortin2024} LMXB catalog lists $339$ Galactic LMXBs and candidates, while XRBcats contains $349$ systems, including black-hole systems and candidates \citep{Avakyan2023}.  Of the XRBcats sources, $179$ have secure NS identifications from thermonuclear bursts or coherent X-ray pulsations, and nine more are classified as NS systems from their X-ray spectra \citep{Niang2024}.  Thus, the observed NS-LMXB sample is already $\gtrsim200$ systems, with new LMXB transients continuing to be discovered regularly \citep{Fortin2024} and additional unclassified LMXBs likely containing NS accretors. This provides the lower limit for $N_{\rm Gal}$.

This lower limit underestimates the intrinsic population.  Many NS-LMXBs are transient and spend most of their time in quiescence, where they would be missed by all-sky X-ray monitoring \citep[e.g.,][and references therein]{Bahramian2023,Heinke2025}. Furthermore, \citet{Maccarone2022} argue that the unseen NS-LMXB population may be much larger than the catalogued sample.  First, they argue that about half of the known transient NS-LMXBs with secure NS accretors have recurrence times longer than a decade for outbursts detectable by all-sky monitors such as MAXI. Second, the Galactic millisecond-pulsar (MSP) population gives an independent birth rate constraint. 
Recycled MSPs are expected to descend from NS-LMXBs (or, less frequently, SyXBs) and live for over a Hubble time. The total Galactic MSP population is about $40{,}000$ \citep{Lorimer2008,Lorimer2013} with a typical age of $\approx5\times10^9\,{\rm yr}$. Thus, it is expected that $N_{\rm NS-LMXB}\approx 40{,}000(\tau_{\rm LMXB}/5\times10^9\,{\rm yr})$.  For $\tau_{\rm LMXB}\gtrsim10^8$~yr, this implies $\gtrsim800$ NS-LMXBs, while $N_{\rm NS-LMXB}\sim200$ would require an X-ray-bright lifetime of only $\sim2\times10^7$ yr \citep{Lorimer2008,Podsiadlowski2002,Maccarone2022}.  
These estimates suggest that the cataloged NS-LMXB sample likely significantly underrepresents the total Galactic NS-LMXB population. The arguments of \citet{Maccarone2022} suggest that there should be $\sim20$ -- $100$ undiscovered NS LMXBs for every one that is known.

Given these large uncertainties, we adopt an intrinsic present-day NS-LMXB population of $N_{\rm Gal} = 300$ -- $30{,}000$, where the lower limit is the size of the known sample and the upper limit is the upper range from \citet{Maccarone2022} to account for the abundance of MSPs.  For a mass-transfer lifetime of order $\tau = 5\times10^8$ -- $10^{10}~{\rm yr}$, this gives a broad birth-rate range of $\mathcal{R} = 3\times10^{-8}$ -- $6\times10^{-5}~{\rm yr^{-1}}$.
For NS-LMXBs, we thus use X-ray-active lifetime $\tau$ and estimated intrinsic number $N_{\rm Gal}$ as inputs to derive $\mathcal{R}$. The $N_{\rm Gal}$ of NS-LMXBs is arguably the least constrained parameter among the NS populations discussed here, owing to uncertainties in recurrence timescales and quiescence properties of the X-ray-selected population. 

The LMXB estimate used here includes known ultracompact X-ray binaries (UCXBs), but we do not estimate their population separately. Population-synthesis calculations predict a large, mostly X-ray-faint UCXB population whose formation rate may exceed that of hydrogen-rich LMXBs \citep{vanHaaften2013,vanHaaften2015}. Their uncertain lifetimes and distinct formation channels prevent a comparably empirical rate estimate, and most proposed UCXB channels do not pass through the detached, au-scale Gaia NS phase.

\subsection{Spider pulsars}\label{subsubsec:rates_spiders}

Spider pulsars are compact binary millisecond pulsars (MSPs) where the recycled NS irradiates and ablates a low-mass companion \citep{Fruchter1988,Roberts2013, Blanchard2025}. They are usually separated into two classes by companion mass: (1) black widows, with very low-mass ($M_c\lesssim0.05$ -- $0.1~{\rm M_\odot}$), often degenerate brown dwarf or white dwarf companions, and (2) redbacks, with more massive, often non-degenerate companions \citep[$M_c\sim0.1$ -- $0.5~{\rm M_\odot}$;][]{Roberts2013,Ginzburg2021,Koljonen2025}. 
SpiderCat lists $111$ Galactic-field spiders and candidates: $50$ confirmed black widows, $30$ confirmed redbacks, five black-widow candidates, $23$ redback candidates, and three rarer systems/candidates \citep{Koljonen2025}.  This observed count is a lower limit of $N_{\rm Gal}$ because radio eclipses, beaming, optical follow-up, and Galactic-plane selection all bias the observations. A significant fraction of the known spider sample resides in globular clusters\footnote{e.g., see \url{https://www3.mpifr-bonn.mpg.de/staff/pfreire/GCpsr.html}}, which we choose to include in our rate estimations because they nominally reside in the Milky Way.
Note that this provides an additional bias because dense-cluster dynamics may enhance the spider formation rate, and cluster pulsars have also been specifically targeted by deeper and longer-baseline radio searches \citep[e.g.,][]{Camilo2005,Ransom2008,Bahramian2013,Verbunt2014,Ye2019,Pan2021,Douglas2022,Gautam2022,Padmanabh2024}.

A direct, spider-only Galactic completeness correction is still uncertain, so we choose to normalize them to the broader MSP population.  The total Galactic MSP population is expected to contain $\sim30{,}000-40{,}000$ sources \citep[see also][]{Lorimer2008,Swiggum2014}.  If spiders constitute $\sim1$ -- $20\%$ of the Galactic MSP population, the implied intrinsic spider population is $N_{\rm Gal}\sim3\times10^2$ -- $6\times10^3$. 
Modeling pulsar irradiation slows the companion's cooling and gives Gyr-scale spider lifetimes, with a short-lived redback phase of $\sim1$ Gyr compared to the longer black widow phase of several Gyr \citep{Ginzburg2020,Ginzburg2021, ConradBurton2023}. 
We therefore adopt a broad range of $\tau_{\rm spider}=1$ -- $10$ Gyr \citep[see also][]{Romani2016}, with a fiducial value of $3$ Gyr, providing a rate $\mathcal{R}=3\times10^{-8}$ -- $6\times10^{-6}~{\rm yr^{-1}}$.

The spider-binary birth rate is also a useful lower bound on the true NS-LMXB formation rate.  If spiders are produced through recycled NS-LMXB evolution, then every spider requires a previous NS-LMXB phase, while only a subset of NS-LMXBs should evolve into observable black widows or redbacks: i.e., $\mathcal{R}_{\rm spider}\leq\mathcal{R}_{\rm NS-LMXB}$. The outcome depends also on the orbital period at the onset of mass transfer, where a bifurcation near $\sim10$--$20~{\rm hr}$ based on the evolutionary state of the donor separates shortening systems that can form spiders or compact PSR--WD binaries from widening systems that form wider PSR--WD binaries \citep{Podsiadlowski2002,Istrate2014,Bellm2016}.

Comparing their current {\it observed} rates provides additional support that the observed NS-LMXB sample likely underestimates the underlying parent population \citep{KulkarniNarayan1988}.

\subsection{Recycled pulsar--WD binaries}\label{subsubsec:rates_psrWD}
White dwarfs (WDs) are the most commonly observed companions to MSPs \citep[e.g.,][]{Kilic2018}.  
These binaries are the end products of mass transfer onto a NS accretor: LMXB progenitors with sufficiently evolved low-mass donors generally leave He-WD companions after they are stripped, while intermediate-mass X-ray binary progenitors can leave more massive CO or ONeMg WDs \citep{Camilo1996,Bhattacharya1991,Rappaport1995,Tauris2000,Podsiadlowski2002,Lorimer2008,Tauris2012}. He-WD companions dominate the binary MSP population, with typical companion masses $M_{\rm WD}\approx0.15$ -- $0.45~{\rm M_\odot}$ and final orbital periods from $<1$~day to $\sim100$~days, following the standard $M_{\rm WD}$ -- $P_{\rm orb}$ relation \citep{Rappaport1995}.

The detached Gaia NS binaries provide a related, but probably subdominant, route to forming PSR--WD binaries.  
\citet{Chattopadhyay2026} evolve the Gaia NS binaries and find that all terminate as wide NS--WD systems through stable mass transfer.
The degree of recycling depends on the treatment of mass transfer: models that assume circularization before Roche-lobe overflow produce MSPs, while models that allow eccentric, periastron-concentrated mass transfer produce only mildly recycled pulsars \citep{Chattopadhyay2026}.
All of their final PSR--WD binaries have $P_{\rm orb}\gtrsim200$~days. Most of the $\sim200$ known MSP--WD binaries outside globular clusters occupy shorter-period, nearly circular orbits, with $P_{\rm orb}<100$~days \citep{Bhattacharya1991,Kiziltan2010,Chattopadhyay2026}, although a small long-period tail of mildly recycled PSR--WD systems is known at $P_{\rm orb}\sim500$ -- $1000$~d and likely descended from Gaia NSs \citep[e.g.,][]{Stovall2014,Lorimer2005,Lorimer2006,Knispel2013,Gonzalez2011,Martinez2019,Wang2025,Yang2026,Tan2026}.

These results suggest that Gaia NSs are likely a subdominant formation pathway for the full Galactic PSR--WD population.
The dominant channel is more likely the standard recycled-pulsar pathway through NS LMXBs or IMXBs with evolved donors, which naturally produces MSPs with He-, CO-, or ONeMg-WD companions over the observed range of compact binary periods \citep{Bhattacharya1991,Rappaport1995,Podsiadlowski2002,Tauris2012,Istrate2014}.
Additional compact MSP--WD systems may also descend from Gaia NSs with somewhat more massive MS donors that undergo common-envelope evolution with the NS and shrink, as inferred for Cyg X-2-like systems \citep{Podsiadlowski2000, Podsiadlowski2002,Tauris2012,Chattopadhyay2026}.

The observed PSR--WD sample is radio-selected and therefore incomplete because of beaming, flux limits, sky coverage, interstellar dispersion, and orbital-acceleration smearing in short-period binaries \citep{Ransom2003}.
We therefore estimate the intrinsic number by scaling from the completeness-corrected Galactic MSP population.  
\citet{Lorimer2013} infer $\sim30{,}000-40{,}000$ Galactic MSPs, which is similar to the less-certain value estimated from Galactic $\gamma$-ray-selected MSPs of $\sim10^4$ \citep[][]{Gregoire2013}.
Because WD companions dominate the ordinary binary MSP population, we adopt a broad WD-companion fraction $f_{\rm WD}\approx0.5$ -- $0.9$. 
This provides
\begin{equation}
    N_{\rm Gal,MSP-WD} = f_{\rm WD} N_{\rm Gal,MSP}
    \approx (1.5-4)\times10^4 ,
\end{equation}
for $N_{\rm Gal,MSP}=30{,}000$ -- $40{,}000$.  
We therefore adopt $N_{\rm Gal,MSP-WD}\sim30{,}000$ recycled pulsar--WD systems, with an order-of-magnitude range $10^4$ -- $5\times10^4$ to reflect the radio-selection and companion-class uncertainties. 
Given the long spin-down lifetimes of recycled MSPs, typically $\sim10^9$ -- $10^{10}~{\rm yr}$, the classical MSP birthrate inferred from the Galactic sample is at least a few $\times10^{-6}~{\rm yr^{-1}}$ \citep{Lorimer2001, Sautron2026}. We adopt $\mathcal{R}=1.5\times10^{-6}$ -- $5\times10^{-5}~{\rm yr^{-1}}$ for the recycled pulsar--WD population, with a fiducial $\mathcal{R}=8\times10^{-6}~{\rm yr^{-1}}$.
For recycled pulsar--WD, we thus use the intrinsic number and recycled-pulsar lifetime as inputs to derive $\mathcal{R}$.

\subsection{Double neutron stars}\label{subsubsec:rates_doubleNS}

The Galactic double neutron star (DNS) population provides one of the best empirical constraints on the formation of compact NS binaries.  Most previous work has focused on the subset that will merge within a Hubble time, because those systems provide an observational anchor for binary neutron-star merger rates from gravitational wave detectors.  Here, however, we are interested in estimating the present-day Galactic DNS population as a whole, including wide systems with merger times longer than the age of the Universe.

Galactic DNSs are identified through radio pulsar timing, so the observed sample is strongly affected by radio-selection effects, including beaming, luminosity limits, sky coverage, interstellar dispersion and scattering, and reduced sensitivity to highly accelerated compact binaries \citep{Ransom2003,Lorimer2008}.  The total DNS population is therefore less directly measured than the DNS merger rate.  Nevertheless, the nearby radio sample gives a useful observational lower limit.  Several long-lived DNSs are known within $\sim1$ kpc: PSR~J1518+4904 has $P_{\rm orb}=8.63$ d and $d=625$ pc, PSR~J0453+1559 has $P_{\rm orb}=4.07$ d and a gravitational-wave merger time $1.43 \times 10^{12}$~yr, and PSR~J1411+2551 and PSR~J1829+2456 have $P_{\rm orb}=2.62$ and $1.18$ d at distances of $\sim1$ kpc \citep{Janssen2008,Martinez2015,Martinez2017,Haniewicz2021,Ding2024}.  Four such systems within $\sim1.1$~kpc imply an observed local density $\sim0.7~{\rm kpc^{-3}}$, before correcting for radio selection effects.  For an effective nearby detection efficiency of order $1$ -- $10\%$ \citep[e.g.,][]{Osowski2011, Chattopadhyay2020}, this corresponds to an intrinsic DNS density of $\sim7$ -- $70~{\rm kpc^{-3}}$, or $N_{\rm Gal,DNS}\sim3\times10^3$ -- $3\times10^4$ for the effective volume of the Galactic thin disk.  We adopt $N_{\rm Gal,DNS}\sim10^4$ as an observationally anchored fiducial value. This is consistent with the population-synthesis estimate of \citet{Chattopadhyay2020}, who find $N_{\rm Gal,~DNS}\sim2.5\times10^4$ present-day Galactic DNSs, only $\sim34\%$ of which merge within a Hubble time in their model \citep[see also][]{VignaGomez2018,Sgalletta2023}.
Note that our estimate excludes even wider DNS binaries ($P \gtrsim100$~days), which are biased against for finite observational baselines.

The rate is better constrained for the merging DNS subset.  Using the observed radio-DNS sample together with pulsar-survey selection functions, \citet{Pol2020} infer a Galactic DNS merger rate of $\mathcal{R}_{\rm MW}=37^{+24}_{-11}~{\rm Myr^{-1}}$.  Updated treatments of the pulsar beaming geometry and the addition of newly discovered systems give consistent values of $\mathcal{R}_{\rm MW}=32^{+19}_{-9}~{\rm Myr^{-1}}$ \citep{Grunthal2021} and $\mathcal{R}_{\rm MW}=25^{+19}_{-9}~{\rm Myr^{-1}}$ \citep{ColomBernadich2023}.  These estimates apply to the merging subset and are somewhat sensitive to small-number statistics, because only $\sim10$ known Galactic DNSs will merge within a Hubble time. 

Given that the full DNS population also contains a long-lived, non-merging reservoir, we treat the adopted $N_{\rm Gal,DNS}$ and a broad effective lifetime to estimate $\mathcal{R}$.  Compact DNSs can merge within $\lesssim10^8$ -- $10^9$ yr, while wide systems such as PSR~J0453+1559 and PSR~J1518+4904 survive for longer than a Hubble time.  We adopt $\tau_{\rm DNS}\sim10^8$ -- $10^{10}$ yr, with a fiducial value of $10^9$ yr.

\subsection{Gravitational-wave mergers}\label{subsubsec:rates_doublecompactmerger}

The Galactic DNS sample gives an empirical DNS merger rate of a few $\times10^{-5}~{\rm yr^{-1}}$ per Milky Way-like galaxy (Section~\ref{subsubsec:rates_doubleNS}).  
Gravitational-wave detections provide an independent constraint on the same quantity, measured as a local volumetric merger-rate density.  
GWTC-4.0 provides a method-marginalized local BNS merger-rate density of $7.6$ -- $250~{\rm Gpc^{-3}~yr^{-1}}$ \citep{LIGOScientificCollaboration2025}.  
Converting this rate to a Milky-Way-equivalent rate using $n_{\rm MWEG}=0.0116~{\rm Mpc^{-3}}$ \citep{Kopparapu2008,Belczynski2016} gives
\begin{equation}
    \mathcal{R}_{\rm BNS,MW}^{\rm LVK}
    =
    \frac{\mathcal{R}_{\rm BNS}}{n_{\rm MWEG}}
    \approx
    6.6\times10^{-7}
    -
    2.2\times10^{-5}~{\rm yr^{-1}} .
\end{equation}
The upper end of this interval is consistent with the central values inferred from Galactic DNSs.

GWTC-4.0 also gives a local NSBH merger-rate density of $9.1$ -- $84~{\rm Gpc^{-3}~yr^{-1}}$, corresponding to
$\mathcal{R}_{\rm NSBH,MW}^{\rm LVK}\approx8\times10^{-7}$ -- $7\times10^{-6}~{\rm yr^{-1}}$
under the same conversion \citep{LIGOScientificCollaboration2025}.  
These gravitational-wave rates are useful checks on compact-object merger channels, but we do not include them when comparing Galactic NS populations in Figure \ref{fig:ns_population_lifetimes} because they constrain merger rates rather than present-day populations, and it would be partly redundant with the DNS population.

\bibliography{refs}
\bibliographystyle{aasjournal}

\end{document}